\documentclass[aps,prd,twocolumn,showpacs,10pt,superscriptaddress,preprintnumbers,nofootinbib]{revtex4-1}
\pdfoutput=1
\usepackage[english]{babel}
\usepackage{amsmath}
\usepackage{graphicx}
\usepackage{bm}
\usepackage{slashed}

\usepackage{xspace}
\usepackage{hyperref}
\usepackage[usenames,dvipsnames]{xcolor}

\allowdisplaybreaks[4]

\newcommand{\beq}{\begin{equation}\begin{aligned}}
\newcommand{\eeq}{\end{aligned}\end{equation}}

\def\Li{{\rm Li}}

\def\cM{{\mathcal M}}

\def\re{\text{Re}}
\def\im{\text{Im}}

\newcommand{\nn}{\nonumber}

\definecolor{darkyellow}{rgb}{0.5, 0.5, 0.0}
\definecolor{darkpurple}{rgb}{0.5, 0.2, 0.8}
\definecolor{darkblue}{rgb}{0.0, 0.0, 0.8}
\definecolor{darkgreen}{rgb}{0.0, 0.4, 0.0}
\definecolor{darkred}{rgb}{0.5, 0.0, 0.0}

\hypersetup{
    linktocpage,
     colorlinks,
     citecolor=darkgreen,
     linkcolor= darkgreen,
     urlcolor=darkgreen
}

\begin{document}

\preprint{ZU-TH 08/24}

\title{Three-point Energy Correlators in Hadronic Higgs Decays}

\author{Tong-Zhi Yang,}
\email{toyang@physik.uzh.ch}
\affiliation{Physik-Institut, Universit\"at Z\"urich, Winterthurerstrasse 190, CH-8057 Z\"urich, Switzerland}
\author{Xiaoyuan Zhang}
\email{xiaoyuanzhang@g.harvard.edu}
\affiliation{Department of Physics, Harvard University, Cambridge, MA 02138, USA}

\begin{abstract}
    We present the analytic calculation of the leading order three-point energy correlator (EEEC) in hadronic Higgs decays, including both gluon-initiated channel $H\rightarrow g g+X$ and quark-initiated channel $H\rightarrow q\bar q+X$. The phase space integration is evaluated directly using Mandelstam variables $s_{ij}=(p_i+p_j)^2$, and the appearing square roots can be rationalized by either conformal ratios or celestial coordinate variables. Throughout the calculation, we observe the same transcendental function space as in $\mathcal{N}=4$ super Yang-Mills (SYM) theory and $e^+e^-\rightarrow \text{ hadrons}$. Different infrared limits are also explored using the full analytic result, offering the fixed-order data for EEEC factorization and resummation. Given its non-trivial shape dependence, the EEEC presents an excellent opportunity to explore the dynamics of gluon jets originating from the $H \to gg$ decay channel at future lepton colliders.
\end{abstract}

\maketitle
\phantom{x}

\section{Introduction}
\label{sec:introduction}

The discovery of the Higgs boson in 2012 by the Large Hadron Collider (LHC)~\cite{ATLAS:2012yve,CMS:2012qbp} has filled in the last piece of the standard model and has made the precise measurement of its properties a primary task in particle physics. This includes determining the Higgs boson mass and its couplings with standard model particles. On one hand, there have been lots of successful progress in understanding the Higgs productions~\cite{Cepeda:2019klc}, in particular, via gluon fusion~\cite{Anastasiou:2002yz,Ravindran:2003um,Anastasiou:2015vya,Anastasiou:2016cez,Mistlberger:2018etf,Chen:2021isd}. On the other hand, due to the large QCD backgrounds at the LHC and high luminosity LHC (HL-LHC)~\cite{ZurbanoFernandez:2020cco}, it is harder to precisely measure the Higgs decay channels and probe the Yukawa couplings with fermions of the first two generations.
While the $H\to b \bar{b}$ decay channel was recently observed at the LHC~\cite{ATLAS:2018kot,CMS:2018nsn}, the channel of Higgs decaying into charm quarks remains unobserved. It is also difficult to search for possible invisible decay channels beyond the standard model, which is one of the directions for new physics exploration.
To reach higher accuracy, people have proposed to build a Higgs factory, such as CEPC~\cite{CEPCStudyGroup:2018rmc,CEPCStudyGroup:2018ghi,CEPCStudyGroup:2023quu}, ILC~\cite{Behnke:2013xla,ILC:2013jhg} and FCC-ee~\cite{TLEPDesignStudyWorkingGroup:2013myl}. At lepton colliders, the environment is cleaner so particle reconstruction and process selection are more precise. In particular, the Higgs decay widths can be determined with high accuracy, making it more promising to explore new physics channels.

Among the Higgs decay channels, the hadronic ones are particularly interesting in QCD, where the final-state particles include hadrons initiated by gluons or quarks (for a review, see Ref.~\cite{Spira:2016ztx}). We will refer to them as $Hgg$ and $Hq\bar q$ decay channels respectively. The $Hgg$ decay rate has been calculated to N${}^4$LO~\cite{Herzog:2017dtz,Djouadi:1991tka,Chetyrkin:1997iv,Baikov:2006ch,Moch:2007tx,Davies:2017rle} in Higgs effective field theory (HEFT)~\cite{PhysRevLett.39.1304,Shifman:1978zn,Inami:1982xt} with top quarks integrated out. The $Hb\bar b$ decay rate in the massless quark approximation is also known to N${}^4$LO~\cite{Herzog:2017dtz,Baikov:2005rw,Braaten:1980yq,Gorishnii:1990zu,Chetyrkin:1996sr} and the bottom mass correction estimated by an effective Higgs-bottom interaction is calculated to N${}^4$LO~\cite{Chetyrkin:1997vj,Davies:2017xsp}.

In this work, we mainly focus on one of the event shape observables, {\it energy correlators}, and apply it to hadronic Higgs decays. Energy correlators measure the energies deposited in the detectors as a function of the angles among these detectors. The two-point case, also known as the energy-energy correlation (EEC) function, was proposed in 1970s~\cite{Basham:1978zq,Basham:1978bw} and has been generalized to higher points recently~\cite{Chen:2019bpb,Chen:2020vvp}. In perturbative QCD, the $n$-point energy correlator is defined as a weighted cross-section
\begin{align}\label{eq:def}
     \frac{d\sigma}{dx_{12}\cdots dx_{(n-1)n}} &\equiv \sum_{m}\sum_{1\leq i_1,\cdots i_n \leq m}\int d\sigma_{m}\nn\\
     &\hspace{-1.5cm}\times  \prod_{1\leq k \leq n}\frac{E_{i_k}}{Q} \prod_{1\leq j < l \leq n}\delta\left(x_{jl}-\frac{1-\cos\theta_{i_{j}i_{l}}}{2}\right)\,,
\end{align}
where $m$ is the number of final-state particles and $d\sigma_m$ is the associated differential cross section. We will utilize $x_{jl}$ to represent the angular distance in the following. Energy correlator is {\it perhaps} the simplest event shape observable to compute analytically. The two-point case, EEC, which depends on one angular variable $x_{12}$ only, has been calculated analytically to NNLO in $\mathcal{N}=4$ SYM~\cite{Belitsky:2013ofa,Henn:2019gkr}, and to NLO in both $e^+ e^-$ annihilation and hadronic Higgs decays~\cite{Dixon:2018qgp,Luo:2019nig,Gao:2020vyx}. For the three-point energy correlator (EEEC) which depends on three angular variables $x_{12},\,x_{13},\,x_{23}$, the LO corresponds to the tree-level four-particle final states if we ignore self-correlation. Due to the soft suppression from energy weights and collinear regularization by the angles, the tree level itself is infrared finite. In Refs.~\cite{Yan:2022cye,Yang:2022tgm}, the LO EEEC was computed in both $\mathcal{N}=4$ SYM and $e^+ e^-$ collision, and the results exhibit simple structure but rich physics. Very recently, the collinear limit of the four-point energy correlator (EEEEC) was calculated in $\mathcal{N}=4$ SYM~\cite{Chicherin:2024ifn}. These advancements strongly motivate us to complete the LO EEEC calculation in hadronic Higgs decay. In the meantime, there are also available numerical programs for computing event shapes, including EEC, at $e^+e^-$ collisions. Examples include, \textsc{Event2}~\cite{Catani:1996jh,Catani:1996vz} and \textsc{Nlojet++}~\cite{Nagy:2001fj,Nagy:2003tz} at NLO,  \textsc{Eerad3}~\cite{Gehrmann-DeRidder:2014hxk} and \textsc{Colorfulnnlo}~\cite{Somogyi:2006da,Somogyi:2006db,Aglietti:2008fe} at NNLO. Very recently, the numerical results at NLO for event shape observables in hadronic Higgs decays were also available~\cite{Gehrmann-DeRidder:2023uld}.

It has been observed that energy correlators have the potential to be a highly accurate event shape in QCD measurements~\cite{Komiske:2022enw,Neill:2022lqx,Chen:2022swd,Liu:2022wop,Liu:2023aqb,Cao:2023oef,Lee:2022ige,Craft:2022kdo,Devereaux:2023vjz,Kang:2023gvg,Holguin:2022epo,Holguin:2023bjf,Holguin:2023dtd}. On the theoretical side, in addition to the potential for reaching higher fixed-order corrections, energy correlators also feature relatively simple factorization theorems. For EEC, the distribution exhibits singular behavior in both the collinear limit $x_{12}\rightarrow 0$ and the back-to-back limit $x_{12}\rightarrow 1$. The collinear limit undergoes a Dokshitzer-Gribov-Lipatov-Altarelli-Parisi (DGLAP)-type factorization, and the large logarithms $\ln(x_{12})$ have been resumed to the next-to-next-to-leading logarithmic (NNLL) accuracy~\cite{Dixon:2019uzg,Korchemsky:2019nzm}. In the back-to-back limit, a transverse-momentum-dependent (TMD) factorization formula was derived in~\cite{Moult:2018jzp} and the resummation up to N${}^4$LL is already available~\cite{COLLINS1981381,ELLIS198499,deFlorian:2004mp,Tulipant:2017ybb,Moult:2019vou,Ebert:2020sfi,Moult:2022xzt,Duhr:2022yyp}. At hadron colliders, transverse EEC (TEEC) is also studied and resummed to N${}^3$LL~\cite{Gao:2019ojf,Gao:2023ivm}. For EEEC, the factorization in the triple collinear limit ($x_{12},x_{13},x_{23}\rightarrow 0$) takes a similar form as EEC and the resummation has reached NNLL accuracy~\cite{Chen:2020vvp,Chen:2023zlx}. Both TEEC and EEEC have been measured at the LHC and used to extract the strong coupling constant $\alpha_s$~\cite{ATLAS:2017qir,ATLAS:2020mee,ATLAS:2023tgo,CMS:2023wcp}. Moreover, thanks to the advancements in track function calculations~\cite{Li:2021zcf,Jaarsma:2022kdd,Chen:2022muj,Chen:2022pdu,Lee:2023tkr}, one can effectively utilize track information from colliders and combine it with energy correlators. This allows us to track the flavors of QCD events and probe the non-perturbative dynamics within the jets~\cite{Chang:2013rca,Chen:2020vvp,Jaarsma:2023ell,Lee:2023xzv}. On the experimental side, the angle resolution is better for charged particles compared to neutral particles. Therefore, it is also interesting to measure the charged energy correlator~\cite{Chen:2020vvp,Lee:2023npz}, i.e. we only sum over the charged particles in Eq.~\eqref{eq:def}. In summary, energy correlator has the potential to reach the frontier of precision QCD in the short future. 

While the quark jet has been extensively studied at lepton colliders, exemplified by processes like $e^+ e^- \to q \bar{q} \to \text{hadrons}$, the understanding of the gluon jet remains limited. The channel involving the decay of the Higgs into two gluons emerges as a promising candidate for studying gluon jets. However, the observation of the $H \to gg$ decay channel at hadron colliders is challenging due to significant background interference. The promise lies in future $e^+ e^-$ colliders, which have the potential to produce exceptionally clean events of the Higgs decaying into two gluons. For instance, the CEPC is anticipated to generate around 7000 events of $H\to gg$ corresponding to a statistical precision of approximately $1.2\%$ according to Ref.~\cite{An:2018dwb}. This opens a gateway to exploring various aspects, including hadronization, of the gluon jet. The traditional event shape observables for $e^+ e^-$ collisions or Higgs decays, such as thrust~\cite{Brandt:1964sa,Farhi:1977sg}, C-parameter~\cite{Parisi:1978eg,Donoghue:1979vi}, heavy jet mass~\cite{Clavelli:1981yh}, and EEC~\cite{Basham:1978zq,Basham:1978bw}, typically depend on a single variable. In contrast, the EEEC, with its nontrivial shape dependence, offers an excellent opportunity to scrutinize the detailed dynamics of gluon jets at future lepton colliders. In this paper, we initialize the studies of EEEC in the context of Higgs physics by first providing the fixed-order data in the hadronic channel.

An outline of the paper is as follows. In Section~\ref{sec:calculation}, we describe the analytic calculation in detail, including the direct integration over the four-particle phase space and the simplification of the obtained results. In section~\ref{sec:results}, we discuss the function space and present the analytic results. 
Notice that EEEC is a function of three angular variables, so we also investigate how to visualize the distribution. In section~\ref{sec:kinematics}, we study the EEEC kinematic limits. With the full analytic result in hand, we can extract the expansion in different infrared limits. This provides the fixed-order data for understanding EEEC factorization and resummation. We conclude in Section~\ref{sec:conclusion}.

\section{Calculation}
\label{sec:calculation}

In this section, we discuss the details of EEEC calculation. First of all, let us write down the relevant interactions of HEFT Lagrangian, where the top quarks are integrated out
\begin{equation}
    \mathcal{L}_{\text{eff}}\supset-\frac{1}{4}\lambda(\mu) \underbrace{H \text{Tr}\left(G^{\mu\nu} G_{\mu\nu} \right)}_{\mathcal{O}_g}+\sum_{q} \frac{y_q(\mu)}{\sqrt{2}} \underbrace{H \bar \psi_q \psi_q}_{\mathcal{O}_q}\,.
\end{equation}
Here $H$ and $\psi_q$ stand for Higgs and quark fields respectively, and $ G^{\mu\nu}$ is the gluon field strength tensor. $\lambda(\mu)$ is the Wilson coefficient with respect to the effective $\mathcal{O}_g$ operator and $y_q(\mu)$ is simply the Yukawa coupling. Note that we also take the massless quark limit (except for the top quark) while keeping the Yukawa coupling non-zero. As discussed in Ref.~\cite{Gao:2019mlt}, there is no interference between $\mathcal{O}_g$ and $\mathcal{O}_q$ when calculating the squared matrix elements under the massless quark limit, which remains true to all orders in $\alpha_s$ because of chiral symmetry\footnote{If consider the bottom mass effect, the operator mixing can not be neglected.}. This allows us to compute the gluonic channel and quark channel separately. 

In the following, we will denote the EEEC for gluon-initiated Higgs decay as $H_{gg}(x_1,x_2,x_3)$ and EEEC for quark-initiated decay as $H_{q\bar q}(x_1,x_2,x_3)$, with $x_{1,2,3}$ being the angular distance among three detectors. Concerning normalization, we will scale the differential distribution by the decay width at the Born level, i.e. 
\begin{equation}
    \Gamma_{Hgg}=\frac{1}{64\pi}N_A \lambda(\mu)^2 m_H^3,\quad \Gamma_{Hq\bar q}=\frac{1}{16\pi}C_A y_q^2(\mu) m_H\,,
\end{equation}
where $m_H$ is the mass of Higgs boson, $N_A$ denotes the number of adjoint representation, and $C_A$ is the Casimir invariant with $N_A =8,\, C_A =3$ in QCD.
Then the explicit expression for LO EEEC becomes
\begin{align}
    H_a(x_1,&x_2,x_3)=\frac{1}{\Gamma_{Ha}}\sum_{i,j,k}\int \frac{dPS_{4}}{2 Q}\,|\mathcal{M}_{H\to 4}|^2\,\frac{E_i E_j E_k}{Q^3}\nn\\
    &\times \delta\left(x_1-\frac{1-\cos\theta_{ij}}{2}\right) \delta\left(x_2-\frac{1-\cos\theta_{ik}}{2}\right)\nn\\
    &\times\delta\left(x_3-\frac{1-\cos\theta_{jk}}{2}\right)\,,\quad  a=gg,\, q\bar q
    \label{eq:higgs_eeec}
\end{align}
with $i,j,k$ run over all final-state particles. Here, the center of mass energy $Q=m_H$, $dPS_4$ represents the four-particle phase space measure, and $|\mathcal{M}_{H\to 4}|^2$ denotes the tree-level squared matrix elements for Higgs decaying into four massless partons.

\subsection{Matrix elements}

We calculate the squared matrix elements using the programs \textsc{Qgraf}~\cite{NOGUEIRA1993279} and \textsc{Form}~\cite{Vermaseren:2000nd,Kuipers:2012rf,Ruijl:2017dtg}, and evaluate the color algebra with \textsc{Color} package~\cite{vanRitbergen:1998pn}. For gluon-initiated decay, the born process is $H\to g g$, and the relevant four-parton subprocesses for LO EEEC are 
\begin{align}
    H&\to g(p_1)\, g(p_2)\, g(p_3)\, g(p_4)\,,\nn\\
    H&\to g(p_1)\, g(p_2)\, q(p_3)\, \bar q(p_4)\,,\nn\\
    H&\to q(p_1)\, \bar q(p_2)\, q(p_3)\, \bar q(p_4)\,,
\end{align}
where the last one includes both non-identical and identical quark contributions. For the quark-initiated channel, we have
\begin{align}
    H&\to q(p_1)\, \bar q(p_2)\, g(p_3)\, g(p_4) \,,\nn\\
    H&\to q(p_1)\, \bar q(p_2)\, q(p_3)\, \bar q(p_4) \,. 
\end{align}
\begin{figure}[!htbp]
    \centering
    \includegraphics[width=0.45\textwidth]{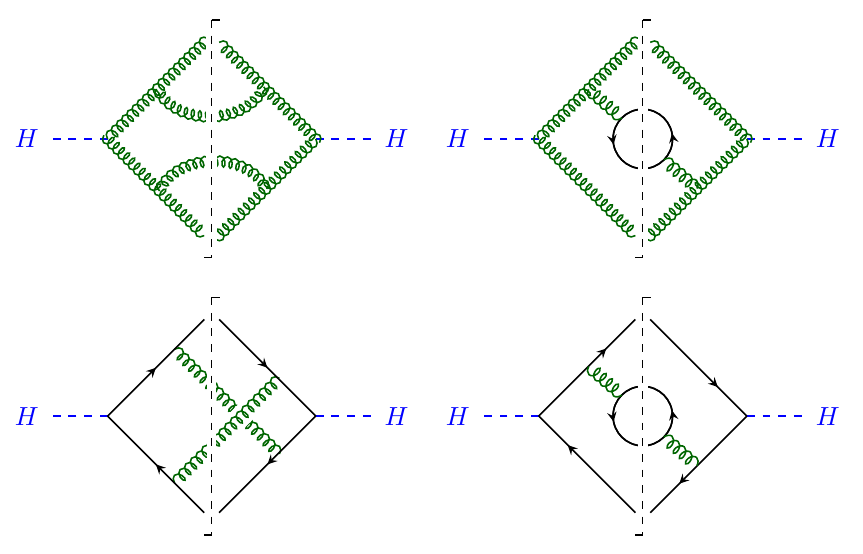}
    \caption{Some example Feynman diagrams for hadronic Higgs decay at $\mathcal{O}(\alpha_s^2)$. The two diagrams on the left represent the $H\rightarrow g g +X$ channel and the other two are the $H\rightarrow q\bar q+X$ channel.}
    \label{fig:HiggsFeynDiag}
\end{figure}

In Fig.~\ref{fig:HiggsFeynDiag}, we present some sample Feynman diagrams.
For all processes, we use the axial gauge for gluon polarization,
\begin{equation}
    \sum_{\lambda=1}^2 \epsilon^\mu (p_i,\lambda)\epsilon^{\ast \nu}(p_i,\lambda)=-g^{\mu\nu} +\frac{\bar n^\mu p_i^\nu+\bar n^\nu p_i^\mu}{\bar n\cdot p_i}-\frac{\bar n^2 p_i^\mu p_i^\nu}{(p_i\cdot \bar n)^2}\,, 
\end{equation}
where $\bar{n}$ is a reference vector. In practice, we choose $\bar{n}$ to be $p_j$ with $i\neq j$ during the summation of the gluon polarization with momentum $p_i$. Alternatively, we can also carry out the calculation in the Feynman gauge where contributions from ghost diagrams are necessary. In the case of the process with four-gluon final states, we need to consider the following two additional processes and their permutations:    
\begin{align}
    &H \to g(p_1) g(p_2) c(p_3) \bar{c}(p_4)\,, \nonumber \\
    & H \to c(p_1) \bar{c}(p_2) c(p_3) \bar{c}(p_4) \,,
\end{align} 
where $c$ stands for ghost particle. We verify that the results obtained in the axial and Feynman gauge are identical. Eventually, we can express the squared matrix elements in terms of Mandelstam variables $s_{ij}=(p_i+p_j)^2$.

To simplify the calculation, we first apply the topology identification. For standard Feynman integrals, renaming the final-state momenta or shifting the loop momenta still leads to the same result. This symmetry can be used to reduce the matrix elements or amplitudes. As discussed in Ref.~\cite{Yang:2022tgm}, in the context of EEEC, we need to carry the energy weights and the $\delta$ measurement functions throughout the calculation (cf. Eq. (2.3) of \cite{Yang:2022tgm}). Eventually, 
we can rewrite Eq.~\eqref{eq:higgs_eeec} as:
\begin{align}
    H_a(&x_1,x_2,x_3)=\frac{1}{\Gamma_{Ha}} \int \frac{dPS_4}{2 Q} |\mathcal{M}_{\text{tot}}(p_1,p_2,p_3,p_4)|^2 \nn\\
    &\times \frac{E_1 E_2 E_3}{Q^3} \delta\left(x_1-\frac{1-\cos\theta_{23}}{2}\right) \delta\left(x_2-\frac{1-\cos\theta_{13}}{2}\right)\nn\\
    &\times\delta\left(x_3-\frac{1-\cos\theta_{12}}{2}\right)+\text{permutations of }x_1,\, x_2,\, x_3 ,\,
\end{align}
with the combined matrix elements
\begin{align}
    |\mathcal{M}_{\text{tot}}(&p_1,p_2,p_3,p_4)|^2=\nn\\
    \big(&|\cM(p_1,p_2,p_3,p_4)|^2+|\cM(p_1,p_2,p_4,p_3)|^2\notag\\
    +&|\cM(p_1,p_4,p_3,p_2)|^2+|\cM(p_4,p_1,p_2,p_3)|^2\big),
\end{align}
where $|\mathcal{M}{(p_1,p_2,p_3,p_4)}|^2 = |\mathcal{M}_{H \to 4}|^2$. Therefore, our task is to compute the integrals for unsymmetric terms and permute the angular distances $x_{1,2,3}$ eventually. 

\subsection{Phase space integration}

The phase space measure for four massless partons can be written as~\cite{Gehrmann-DeRidder:2003pne}
\begin{align}
    {\text dPS}_4&= (2\pi)^{4-3d} (Q^2)^{1-\frac{d}{2}} 2^{1-2d} (-\Delta_4)^{\frac{d-5}{2}}\Theta(-\Delta_4)\nn\\
    &\times d\Omega_{d-1}d\Omega_{d-2}d\Omega_{d-3}ds_{12}ds_{13}ds_{14}ds_{23}ds_{24}ds_{34}\nn\\
    &\times \delta(Q^2-s_{12}-s_{13}-s_{14}-s_{23}-s_{24}-s_{34})\,.
    \label{eq:dps4}
\end{align}
In the direct phase space integration method, the primary challenge arises from intricate constraints involving the summations of high powers of $s_{ij}$ due to the presence of the Heaviside function $\Theta(-\Delta_4)$, where the Gram determinant is defined as $\Delta_4 = \lambda (s_{12}s_{34}, s_{13}s_{24}, s_{14}s_{23})$, with $\lambda$ being the K\"{a}ll\'{e}n function:
\begin{equation}
    \lambda(x, y, z)= x^2+y^2+z^2-2(x y +x z+y z)\,.
\end{equation}
Fortunately, by introducing the energy fractions for the first three particles, defined as 
\begin{align}
    z_i\equiv\frac{2 p_i\cdot q}{Q^2},\,i=1,2,3 \,,
\end{align}
the non-analytic Heaviside step function $\Theta(-\Delta_4)$ decouples from the phase space integral,
\begin{multline}
    \Delta_4=z_1^2 z_2^2 z_3^2 \left(x_1^2+x_2^2+x_3^2 - 2x_1x_2 -2x_1x_3 \right.\\
    \left.-2x_2x_3+4x_1x_2x_3\right)\equiv z_1^2 z_2^2 z_3^2\, \widetilde \Delta_4
    \label{eq:delta4_fac}
\end{multline}
leaving $\widetilde \Delta_4<0$ as a kinematic constraint for the angular distances $x_{1,2,3}$. To achieve the factorization as shown in Eq.~\eqref{eq:delta4_fac}, we have expressed the $s_{ij}$ through the angular distances $x_{1,2,3}$ and momentum fractions $z_{1,2,3}$, for example,
\begin{align}
    x_1= \frac{1-\cos \theta_{23} }{2} = \frac{2 p_2 \cdot p_3 Q^2}{4 p_2\cdot Q p_3\cdot Q} = \frac{s_{23}}{Q^2 z_2 z_3}\,.
\end{align}

Following the procedure in~\cite{Yang:2022tgm}, we express the EEEC as two-fold integrals over energy fractions $z_1$ and $z_3$:
\begin{align}
\label{eq:integrals}
    \int_0^1 dz_3 \int_0^{\frac{1-z_3}{1- x_2 z_3}}\, dz_1 f\left(x_1,x_2,x_3,z_1,z_3\right)\,
\end{align} 
with $f(x_1, x_2,x_3,z_1,z_3)$ being an appropriate integrand.
To calculate the above integrals, we employ partial fractions on all denominators within the integrand $f(x_1, x_2, x_3, z_1, z_3)$, leading to the emergence of three square roots after a direct $z_1$ integration:
\begin{align}
    &s_1\equiv\sqrt{x_1^2+x_2^2+x_3^2 - 2x_1x_2 -2x_1x_3-2x_2x_3}\,,\nn\\
    &s_2\equiv\sqrt{x_1^2+x_2^2+x_3^2 - 2x_1x_2 -2x_1x_3-2x_2x_3+4x_1x_2x_3}\,,\nn\\
    &s_3\equiv\sqrt{s_1^2 z_3^2+2 \left(x_1+x_2-x_3\right) x_3 z_3+x_3^2}\,.
    \label{eq:squareroot}
\end{align}
The first two square roots don't involve the integration variable $z_3$. Introducing a set of variables via $\frac{x_1}{x_3}= z \bar z$, $\frac{x_2}{x_3}= (1-z)(1-\bar z)$ and $x_3=\frac{t^2-(z-\bar z)^2}{4 z \bar z (1-z)(1-\bar z)}$, both square roots can be rationalized,
\begin{align}
    s_1&=x_3(z-\bar z)=\frac{t^2-(z-\bar z)^2}{4z \bar z (1-z)(1-\bar z)}  (z-\bar z)\,,\nn\\
    s_2&=x_3 t=\frac{t^2-(z-\bar z)^2}{4z \bar z (1-z)(1-\bar z)} t \,.
\end{align}
Notice that $t$ is purely imaginary. For the third one, we pass $z_3$ to a new integration variable $y$
\begin{equation}
    z_3\equiv\frac{y x_3}{y^2 x_2-y \left(x_1+x_2-x_3\right)+x_1}\,,
\end{equation}
such that $s_3$ is rationalized
\begin{equation}
    s_3=\frac{x_3^2 \left(x_1-y^2 x_2\right){}^2}{\left(y^2 x_2-y \left(x_1+x_2-x_3\right)+x_1\right){}^2}\,.
\end{equation}
The integration range for the variable $y$ remains within the interval $[0,1]$. Subsequently, we need to further employ partial fractions to the obtained integrand such that all denominators are linear in $y$. Finally, the $y$ integrals can be evaluated analytically using the program \textsc{Hyperint}~\cite{Panzer:2014caa}, and the result is expressed in terms of Goncharov polylogarithms (GPLs)~\cite{goncharov1mpl,Goncharov:1998kja,Borwein:1999js}. The GPLs are defined by iterative integrals:
\begin{equation}\label{eq:gpl_def}
G(a_1,\cdots a_n; x)\equiv\int_0^x \frac{dt}{t-a_1} G(a_2,\cdots a_n; t)\,,
\end{equation}
with
\begin{equation}
G(;x)\equiv1,\quad G(\vec 0_n;x)\equiv\frac{1}{n!}\ln^n (x)\,.
\end{equation}

At this stage, the numerical EEEC distribution can already be evaluated by publicly available programs such as \textsc{Ginac}\footnote{We utilize the program \textsc{PolyLogTools}~\cite{Duhr:2019tlz} as a backend to invoke \textsc{Ginac}.}~\cite{Bauer:2000cp}, \textsc{HandyG}~\cite{Naterop:2019xaf} and \textsc{Fastgpl}~\cite{Wang:2021imw}. 

\subsection{Simplification}

To speed up the numerical evaluation and to understand the analytic structure of EEEC, we need to simplify the result. It is known that GPLs up to transcendentality weight three can be expressed in terms of logarithms and classical polylogarithms $\text{Li}_n(x)$ with $n\leq 3$. So, first of all, we convert all GPLs in the raw result into classical polylogarithms using \textsc{Gtorules}~\cite{Frellesvig:2016ske}. Some adjustments for the arguments in $\ln(x)$ and $\text{Li}_2(x)$ are required to meet the branch prescription in \textsc{Mathematica}. Then we collect the transcendental functions with the same coefficients and construct the raw function space. The rational coefficients can be simplified by the multivariate partial fraction package \textsc{MultivariateApart}~\cite{Heller:2021qkz}. Regarding the raw bases, we apply the transcendentality weight-two identities to simplify the $\text{Li}_2(x)$ functions. These identities can be generated by the well-known five-term identity~\cite{wojtkowiak1996functional}:
\begin{multline}
    \Li_2(x) + \Li_2(y) + \Li_2\left(\frac{1-x}{1-xy}\right) + \Li_2(1-xy)  \\*
    + \Li_2\left(\frac{1-y}{1-xy}\right)= \frac{\pi^2}{2} - \log(x)\log(1-x)  - \log(y)\log(1-y)\\
    - \log\left(\frac{1-x}{1-xy}\right) \log\left(\frac{1-y}{1-xy}\right)\,.
\end{multline}
After that, we add the $x_{1,2,3}$ permutations and repeat this procedure until the expression cannot be further simplified. 

More interestingly, we find that there exists a linear transformation between the simplified function space and the function space of EEEC in $\mathcal{N}=4$ SYM theory~\cite{Yan:2022cye}. We calculate this transformation by evaluating both function spaces with the same numerical point and applying the \textsc{PSLQ} algorithm~\cite{PSLQref,Bailey:1999nv}. Eventually, the Higgs EEEC is represented in terms of the same function space as in $\mathcal{N}=4$. 

\section{Results}
\label{sec:results}

In this section, we describe the analytic result of Higgs EEEC. 

\subsection{Function space}\label{sec:functionspace}

\begin{figure}[!hbtp]
    \centering
    \includegraphics[scale=0.25]{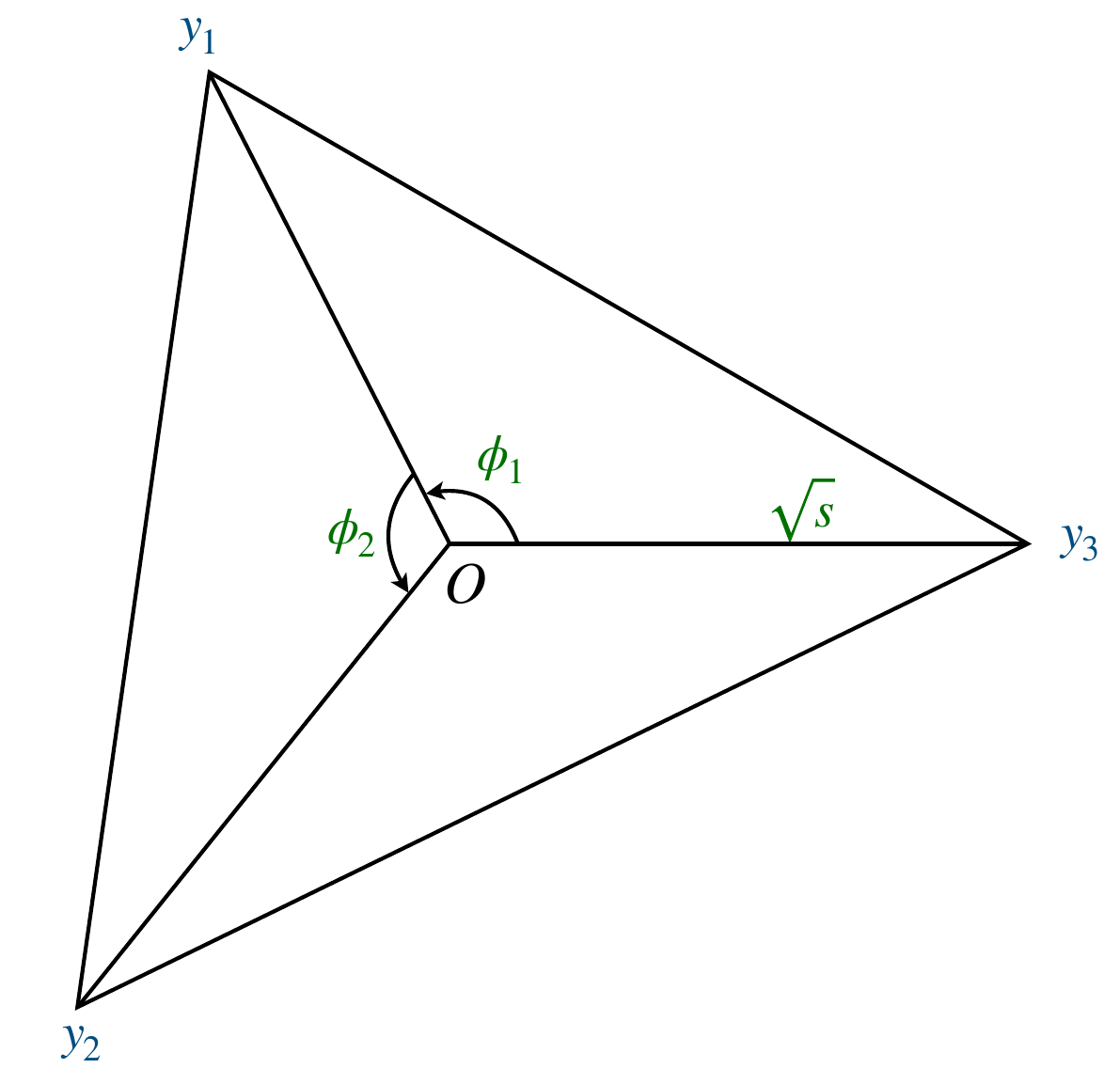}
    \caption{Demonstration of celestial variables. By mapping the angular distances $x_{1,2,3}$ to three points $y_{1,2,3}$ on the celestial sphere, we pass them to the radius $s$ and two angles $\phi_{1,2}$ among these points.}
    \label{fig:ypara}
\end{figure}

The EEEC function space in $\mathcal{N}=4$ SYM was studied in Ref.~\cite{Yan:2022cye}. Following the calculation in Ref.~\cite{Yang:2022tgm} and the previous section, we find $e^+ e^-$ annihilation and hadronic Higgs decays share the same function space. Here we briefly summarize the structure. First of all, we introduce the celestial variables by mapping the angular variables $x_{1,2,3}$ onto the distances among three points $y_{1,2,3}$ on the celestial sphere:
\begin{equation}
    x_i=\frac{|y_j-y_k|^2}{(1+|y_j|^2)(1+|y_k|^2)},\quad i\neq j\neq k\in\{1,2,3\}\,.
    \label{eq:xtoy}
\end{equation}
Since these three points also fall into a circle, we can further map the $y_i$ variables onto the radius $\sqrt{s}$ and two angles $\phi_1$, $\phi_2$ (see Fig.~\ref{fig:ypara}). Explicitly, we define
\begin{equation}
    y_1=\sqrt{s} \underbrace{e^{i\phi_1}}_{\equiv \tau_1},\quad y_2=\sqrt{s} \underbrace{e^{i(\phi_1+\phi_2)}}_{\equiv \tau_1\tau_2},\quad y_3=\sqrt{s}
    \label{eq:ytotau} \,. 
\end{equation}
The $\{s,\tau_1,\tau_2\}$ can also rationalize the first two square roots in Eq.~\eqref{eq:squareroot}. Below we will use this variable set for some cases and the angular distances $\{x_1,x_2,x_3\}$ for others, depending on which one manifests the physics. The relation between them is 
\begin{align}
x_1&=-\frac{s}{(s+1)^2}\frac{(1-\tau_1)^2}{\tau_1},\quad x_2=-\frac{s}{(s+1)^2}\frac{(1-\tau_2)^2}{\tau_2},\nn\\
x_3&=-\frac{s}{(s+1)^2}\frac{(1-\tau_1 \tau_2)^2}{\tau_1 \tau_2}\,.
\label{eq:xtostau}
\end{align}

It is found in Ref.~\cite{Yan:2022cye} that the EEEC kinematic space can be embedded into a hexagon on a unit circle and the remaining symmetry is the dihedral group $D_6$. This implies that EEEC contains some similar structure to six-gluon scattering amplitude. If we define three conformal invariant ratios
\begin{equation}
    u_1\equiv -\frac{s+\tau_1}{1+s\tau_1},\,u_2\equiv \frac{\tau_1-1}{1-\tau_1\tau_2}, \,u_3\equiv \frac{(1-\tau_1)(s+\tau_2)}{(1-\tau_2)(1+s\tau_1)}\,,
\end{equation}
its symbol alphabet can be written as a close set under the $D_6$ group. Finally, incorporating the single-valued requirement and the singularity structure in the triple collinear limit~\cite{Chen:2019bpb}, we are able to reconstruct the entire EEEC function space. The form that manifests these structures can be found in Ref.~\cite{Yan:2022cye}, but here we provide the expression in terms of $x_{1,2,3}$, which is more convenient for phenomenological studies.

The complete function space contains classical polylogarithms up to transcendentality weight two. At weight one, we find seven basis\footnote{When evaluating $\ln\left(\frac{2-s_2-x_1-x_2-x_3}{2+s_2-x_1-x_2-x_3}\right)$ in \textsc{Mathematica}, one need to be careful with the branch cuts. For all allowed values of $x_{1,2,3}$, this logarithm should be rewritten as $\ln\left(2-s_2-x_1-x_2-x_3\right)-\ln\left(2+s_2-x_1-x_2-x_3\right)$.  }
\begin{align}
\label{eq:w1base}
    f_1&=\ln (1-x_1),\quad f_2=\ln x_1, \quad f_3=\ln (1-x_2),\nn\\
    f_4&=\ln x_2, \quad f_5=\ln (1-x_3),f_6=\ln x_3,\nn\\
f_7&=\ln\left(v\right),\quad v=\frac{2-s_2-x_1-x_2-x_3}{2+s_2-x_1-x_2-x_3}\,.
\end{align}
At transcendentality weight two, we find 21 bases, with explicit expressions given in Eq.~\eqref{eq:eeec_tw2_basis} in the appendix. Here we quote a specific basis as an example:
\begin{multline}
\label{eq:g17exp}
    g_{17}=-i\im\left[\text{Li}_2(v)\right]\\
    -\frac{1}{2}\ln(v)\ln\left[\frac{s_2^2}{\left(x_1-1\right) \left(x_2-1\right)
   \left(x_3-1\right)}\right]\,,
\end{multline}
and we emphasize that despite their looking, all bases are analytic within the allowed kinematic space $\widetilde\Delta_4 <0$. That is, the $\im \left[ \text{Li}_2(x) \right]$ in the above equation and in Eq.~\eqref{eq:eeec_tw2_basis} is understood as 
\begin{align}
    \im \left[ \text{Li}_2(x) \right] = \frac{1}{2i}\bigg[ \text{Li}_2(x)- \Big(\text{Li}_2(x)\big|_{s_1 \to -s_1,\,s_2 \to -s_2} \Big)\bigg]\,.
\end{align}

It is surprising that, despite the distinct nature of scattering processes among $\mathcal{N}=4$ SYM, $e^+e^- \to \text{hadrons}$, and hadronic Higgs decay, their EEEC function spaces are exactly the same. This also happens for EEC, as shown in Refs.~\cite{Belitsky:2013ofa,Dixon:2018qgp,Luo:2019nig,Gao:2020vyx}. It will be interesting to understand the universality inside the structure of energy correlators. This might allow us to build a future bootstrap program that can be directly applied to the jet substructures at colliders.

\subsection{Analytic results}

Here we present the analytic result of Higgs EEEC. For gluonic channel $Hgg$, we have
\begin{align}\label{eq:hgg_res_color}
    H_{gg}(&x_1,x_2,x_3)=\left(\frac{\alpha_s}{4\pi}\right)^2\frac{1}{4\pi \sqrt{-s_2^2}}\bigg[T_F^2 n_f^2 A_1(x_i) \nn\\
    &+C_F T_F n_f A_2 (x_i)+ C_A T_F n_f A_3(x_i) +C_A^2\, A_4(x_i) \bigg]\,,
\end{align}
where $A_1(x_i)$ is the simplest term, only containing up to transcendentality weight-one functions $f_{1,3,5,7}$. As an example, we write down $A_1(x_i)$ explicitly:
\begin{widetext}
\begin{align}
    &A_1=\frac{472 x_3^3+1952 x_1 x_2 x_3^2-3004 x_2 x_3^2+1130 x_1^2 x_2^2 x_3-4574 x_1 x_2^2 x_3+2400 x_2^2 x_3+1724 x_1
   x_2 x_3+36 x_1^2 x_2^2-136 x_1 x_2^2}{105 x_1^4 x_2^4 x_3^2}\nn\\
   &+\bigg[\frac{236 x_1^7-1974 x_3 x_1^6+2128 x_2 x_3 x_1^5-3640 x_2^2 x_3^5 x_1^2+6580 x_2 x_3^5 x_1^2-3262 x_3^5
   x_1^2+6580 x_2^2 x_3^5 x_1-4256 x_2 x_3^5 x_1-3262 x_2^2 x_3^5}{105 x_1^5 x_2^5 x_3^5}\nn\\
   &\hspace{0.5cm}+\frac{190 x_2 x_1^3+260 x_2^2 x_3 x_1^2-940 x_2 x_3 x_1^2+466 x_3 x_1^2+1000 x_2 x_3^2 x_1-460 x_3^2 x_1+480
   x_2^2 x_3 x_1-410 x_2 x_3 x_1-40 x_2^2 x_3}{15 x_1^2 x_2^5 x_3^4}\nn\\
   &\hspace{0.5cm}+\frac{20 x_2^3 x_1^3-148 x_2^2 x_1^3+8 x_2^3 x_1^2-200 x_2 x_3 x_1^2+92 x_3 x_1^2-24 x_2^3 x_1+16 x_2^2 x_1-192
   x_2^2 x_3 x_1+82 x_2 x_3 x_1+8 x_2^3+82 x_2^2 x_3}{3 x_1^4 x_2^5 x_3^2}\nn\\
   &\hspace{0.5cm}+\frac{-1400 x_2^3 x_1^3+5180 x_2^2 x_1^3+4620 x_2^3 x_1^2-5320 x_2^3 x_1-2660 x_2 x_3^2 x_1+1974 x_3^2 x_1+2100
   x_2^3-472 x_3^3+1974 x_2 x_3^2}{105 x_1^5 x_2^5 x_3}\bigg]f_1\nn\\
   &+\bigg[ \frac{-2624 x_2 x_3^2 x_1^3+1224 x_2 x_1^3+15544 x_2 x_3^2 x_1^2-9476 x_3^2 x_1^2-976 x_2 x_1^2-17100 x_2 x_3^2
   x_1+7384 x_3^2 x_1+6482 x_2 x_3^2}{105 x_1^5 x_2^3 x_3^2}\nn\\
   &\hspace{0.5cm}+\frac{-240 x_2^2 x_1^4+3568 x_2^2 x_1^3-5112 x_2 x_1^3-8088 x_2^2 x_1^2+7828 x_2 x_1^2-866 x_1^2+8460 x_2^2
   x_1-2570 x_2 x_1-3220 x_2^2}{105 x_1^5 x_2^3 x_3}\nn\\
   &\hspace{0.5cm}+ \frac{-236 x_3^3-1802 x_1 x_2 x_3^2+2446 x_2 x_3^2-4008 x_1^2 x_2^2 x_3+12244 x_1 x_2^2 x_3-5236 x_2^2 x_3-3630
   x_1 x_2 x_3}{105 x_1^5 x_2^5}\bigg] \frac{f_7}{s_2}\nn\\
   &+\text{permutations of } x_1,\, x_2,\, x_3\,.
\end{align}
\end{widetext}
In the ancillary file accompanying this paper, we provide the expressions for other functions $A_{2-4}(x_i)$. For $H q\bar q$, we find
\begin{align}\label{eq:hqq_res_color}
    H_{q\bar q}(x_1,x_2,x_3)&=\left(\frac{\alpha_s}{4\pi}\right)^2\frac{1}{4\pi\sqrt{-s_2^2}}\bigg[C_F T_F n_f B_1(x_i)\nn\\
    &+ C_F^2 B_2(x_i)+C_F C_A B_3(x_i)\bigg]
\end{align}
with explicit forms for each color channel also included in the ancillary file.

\begin{figure}[!htbp]
    \centering
    \includegraphics[width=0.45\textwidth]{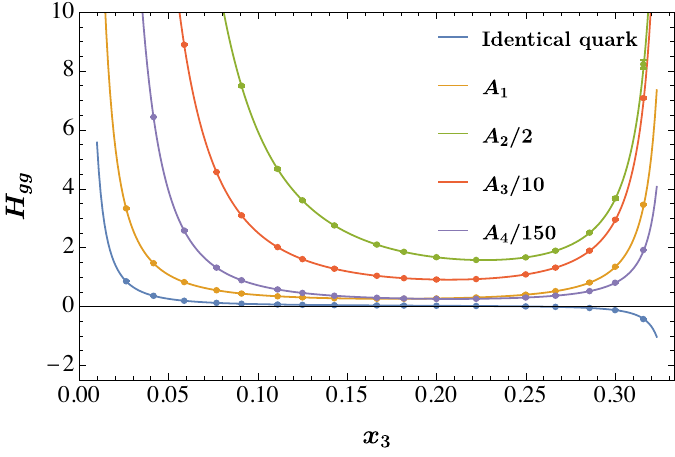}
    \caption{Comparison of our analytic result $H_{gg}$ and numerical evaluation in \textsc{Cuba}. We pick the configuration where the ratios are set as $x_1:x_2:x_3=3:2:1$. $A_{1-4}$ represents different color channels in Eq.~\eqref{eq:hgg_res_color}. The smooth curves are from the analytic result and the points are from \textsc{Cuba}. To show all the curves in the same figure, we rescale the results by some constants. Since the identical quark contribution is tiny, we also verify it separately.}
    \label{fig:hgg_check}
\end{figure}

\begin{figure}[!htbp]
    \centering
    \includegraphics[width=0.45\textwidth]{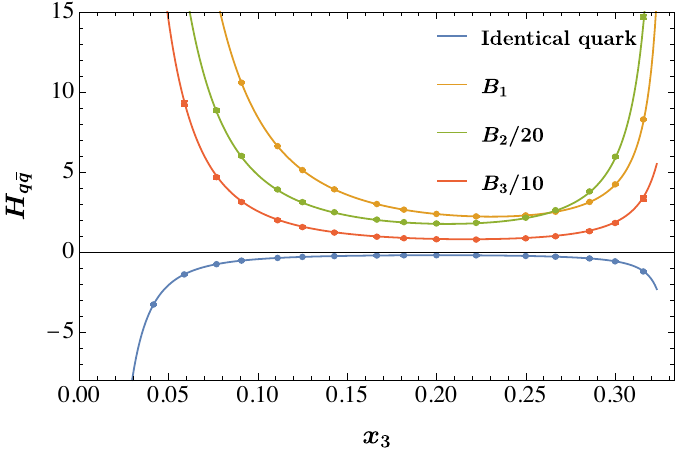}
    \caption{Comparison of our analytic result $H_{q\bar q}$ and numerical evaluation in \textsc{Cuba}. The same configuration is chosen as the gluonic channel. $B_{1-3}$ represents different color channels in Eq.~\eqref{eq:hqq_res_color}. Again, we divide the curves by different constants to make the figure compact.}
    \label{fig:hqq_check}
\end{figure}

To verify the result, we also perform some numerical checks. To our knowledge, there is no publicly available numerical program capable of calculating the fixed-order hadronic Higgs decays\footnote{In \textsc{MadGraph5}~\cite{Alwall:2011uj,Alwall:2014hca}, we find that the \textsc{Heft} model doesn't include the effective vertex $Hgggg$.}. So instead, we generate all matrix elements again in \textsc{FeynCalc}~\cite{Mertig:1990an,Shtabovenko:2016sxi,Shtabovenko:2020gxv,Shtabovenko:2023idz} as a crosscheck and implement the four-particle phase space in Eq.~\eqref{eq:dps4} with the Monte Carlo library \textsc{Cuba}~\cite{Hahn:2004fe}. Since LO EEEC itself is infrared finite, we don't need any subtraction scheme. In Fig.~\ref{fig:hgg_check}, we choose one configuration with ratios $x_1:x_2:x_3=3:2:1$ and calculate 17 points numerically. For each color channel, we find good agreements. Additionally, we also calculate the identical quark pairs contribution separately as a dedicated check. We did the same cross-check for the $Hq\bar q$ EEEC and the comparison can be found in Fig.~\ref{fig:hqq_check}.

We emphasize that simplifying our analytic result into classical polylogarithms makes its numerical evaluation very fast and convenient. In \textsc{Mathematica}, it takes less than 4 seconds to numerically evaluate our analytic formula on a generic point for $H_{gg}$ to 200 digits of accuracy and less than 2 seconds for $H_{q\bar q}$. This is much faster than Monte Carlo simulation. In addition, our result can also provide accurate values for a singular point while it is usually difficult to extract the logarithmic behavior from numerical integration. 

\subsection{Visualization}

Now let us discuss how to visualize the EEEC result. Notice that EEEC is a function of three angular distances, so it is a three-dimensional density distribution. As shown in Eq.~\eqref{eq:delta4_fac}, the kinematic space is constrained by $x_1^2+x_2^2+x_3^2-2x_1x_2-2x_1x_3-2x_2x_3+4x_1x_2x_3<0$, leading to an irregular space (we refer it as `zongzi' space) in Fig.~\ref{fig:kinematic_region}. Although angular distances have greater physical intuition, slicing or projecting an irregular space poses challenges. However, if we instead use $\{s,\phi_1,\phi_2\}$ variables in Eq.~\eqref{eq:xtostau}, the kinematic space becomes a more manageable cube: $s\in [0,1]$ and $\phi_{1,2}/\pi \in [0,1]$.

\begin{figure*}[!htbp]
    \centering
    \includegraphics[scale=0.5]{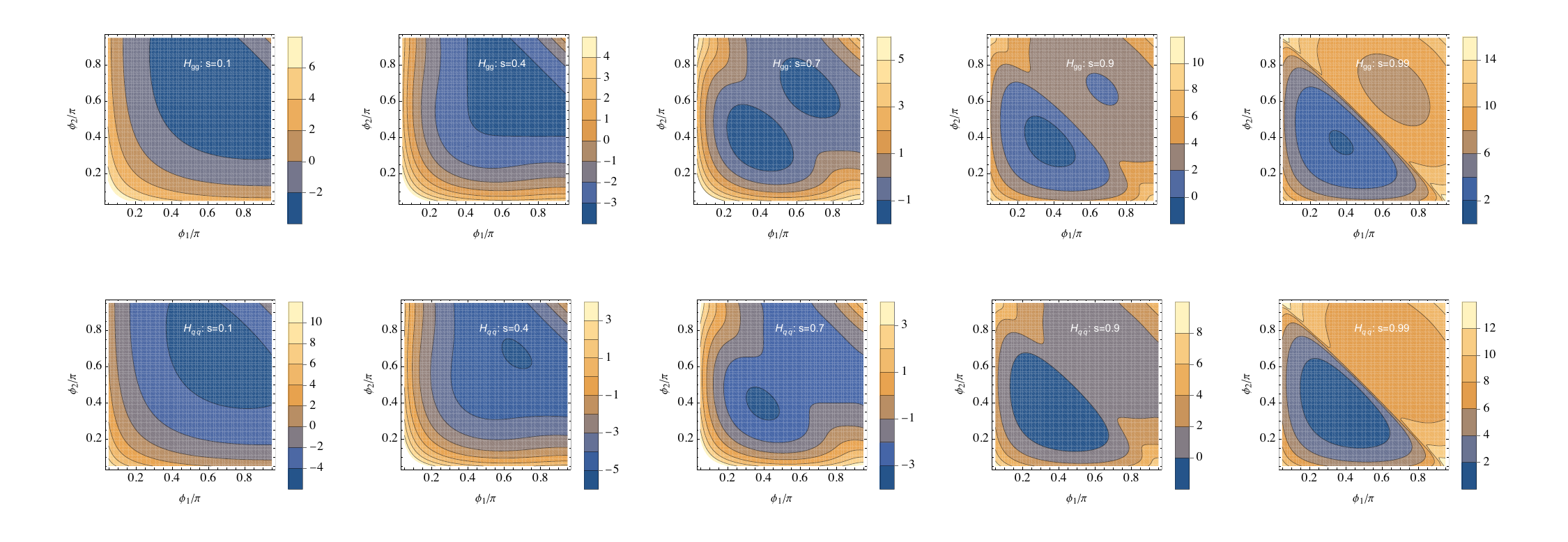}
    \caption{The logarithmic density $\log H_{gg}$ (upper panel) and $\log H_{q\bar q}$ (lower panel) with respect to $\phi_{1,2}$. For a fixed $s$, the distribution gets enhanced when either $\phi_1$ or $\phi_2$ is close to $0$. When $s$ is close to $1$, the distribution receives additional logarithmic enhancement. We set the strong coupling constant $\alpha_s\left(m_H\right)=0.1125$.}
    \label{fig:hgghqqDsLogplot}
\end{figure*}

To begin with, we choose several values for $s$ and plot the logarithmic density $\log H_{gg,q\bar q}$ with respect to $\phi_{1,2}$, as shown in Fig.~\ref{fig:hgghqqDsLogplot}. For every individual figure, we find that the distribution grows fast when $\phi_1\to 0$ or $\phi_2\to 0$. This implies possible large logarithms in this limit. In fact, $\phi_{1}\to 0$ leads to $\tau_{1}\to 1$, thus $x_1\to 0$ and $x_2\approx x_3$, which corresponds to the squeeze limit. The squeeze limit logarithms and the subleading power are studied and resumed using light-ray operator product expansion (OPE) in Refs.~\cite{Chen:2020adz,Chen:2021gdk,Chen:2022jhb,Chen:2023wah,Chen:2023zzh}. There are also additional growing behaviors when $s\to 1$, indicating possible logarithms in the coplanar limit. We will discuss the coplanar limit in more detail next section. Meanwhile, we also observe that $H_{gg}$ and $H_{q\bar q}$ have a similar distribution near the triple collinear limit ($s\to 0$ or $x_{1,2,3}\to 0$) and the coplanar limit, so the gluonic channel and quark channel may share some universal structure in these two limits. 
It will be interesting to understand the difference between $H_{gg}$ and $H_{q\bar q}$ in the middle $s$ region.

\begin{figure}
    \centering
    \includegraphics[scale=0.68]{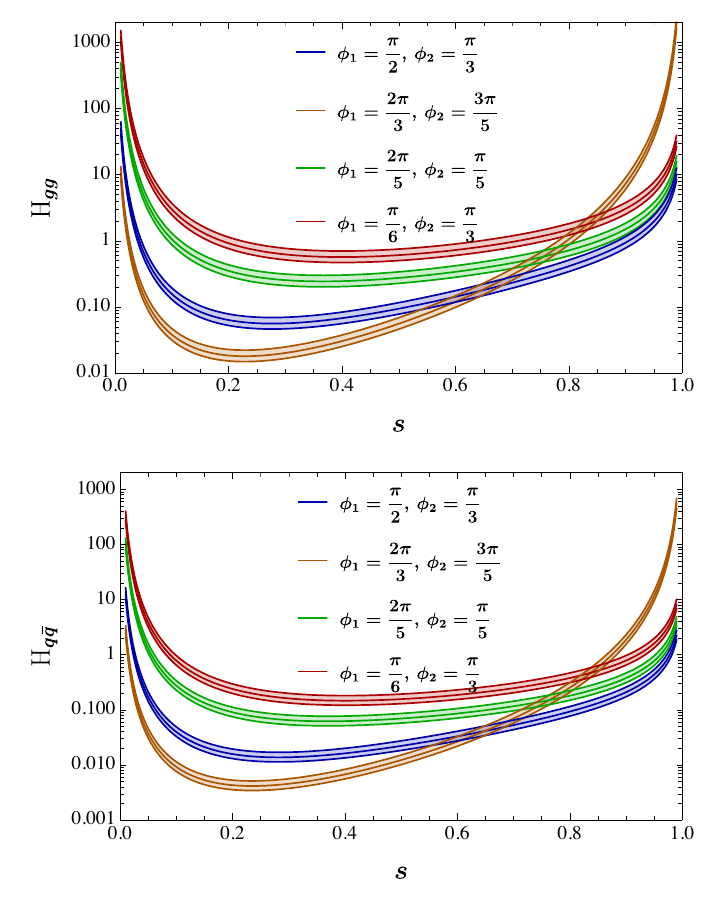}
    \caption{The $s$ dependence of EEEC distribution $H_{gg}$ and $H_{q\bar q}$ with $\phi_1,\phi_2$ fixed. For each channel, we pick four sets of values of $\{\phi_1,\phi_2\}$ for illustration. To estimate uncertainties, we vary the scale $\mu$ in $\alpha_s\left(\mu\right)$ by a factor of 2, both upward and downward from $m_H$.}
    \label{fig:hggqqCsPlot}
\end{figure}

We can also investigate the $s$ dependence of EEEC when fixing the $\phi_{1,2}$. In Fig.~\ref{fig:hggqqCsPlot}, we choose four values $\{\phi_1,\phi_2\}=\{\pi/2,\pi/3\}$, $\{2\pi/3,3\pi/5\}$, $\{2\pi/5,\pi/5\}$, $\{\pi/6,\pi/3\}$ and plot both $H_{gg}$ and $H_{q\bar q}$ EEEC. All curves have similar behaviors as the two-point energy correlator EEC, becoming divergent in both $s\to 0$ and $s\to 1$ edges. The higher-order uncertainty band is estimated by varying the renormalization scale in the strong coupling constant, i.e. $\alpha_s\left(\mu=2^{v}m_H\right),\, v=-1,0,1$.

It is also useful to project EEEC into lower-dimensional distributions. For example, the projected three-point energy correlator is introduced in Ref.~\cite{Chen:2020vvp}, where the longest angular distance $x_L\equiv \text{max}\{x_1,x_2,x_3\}$ is kept and the others are integrated out. This can also be generalized to $\nu$-point correlator~\cite{Chen:2020vvp} as long as $\Re(\nu)>0$, and essentially, we shuffle the information encoded by the kinematic space into different values of $\nu$. To achieve the projected three-point energy correlator for Higgs decay, we will also need to calculate the contact terms, i.e. the differential cross section with energy weight $E_i^2 E_j$ and $E_i^3$ up to NLO. We save this for future work.

\section{Kinematic limits}
\label{sec:kinematics}

\begin{figure}
	\centering
	\includegraphics[scale=0.75]{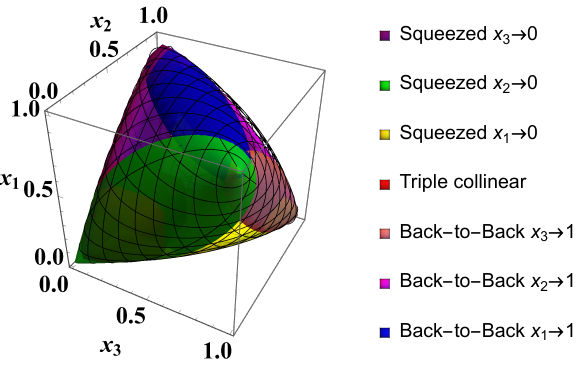}
	\caption{The `zongzi' space. The ranges of $x_{1,2,3}$ allowed by different kinematic limit are shown and the boundary corresponds to the coplanar limit. Notice that there are overlaps among different regions.}
	\label{fig:kinematic_region}
\end{figure}

\begin{figure*}[!htbp]
    \centering
    \includegraphics[scale=0.42]{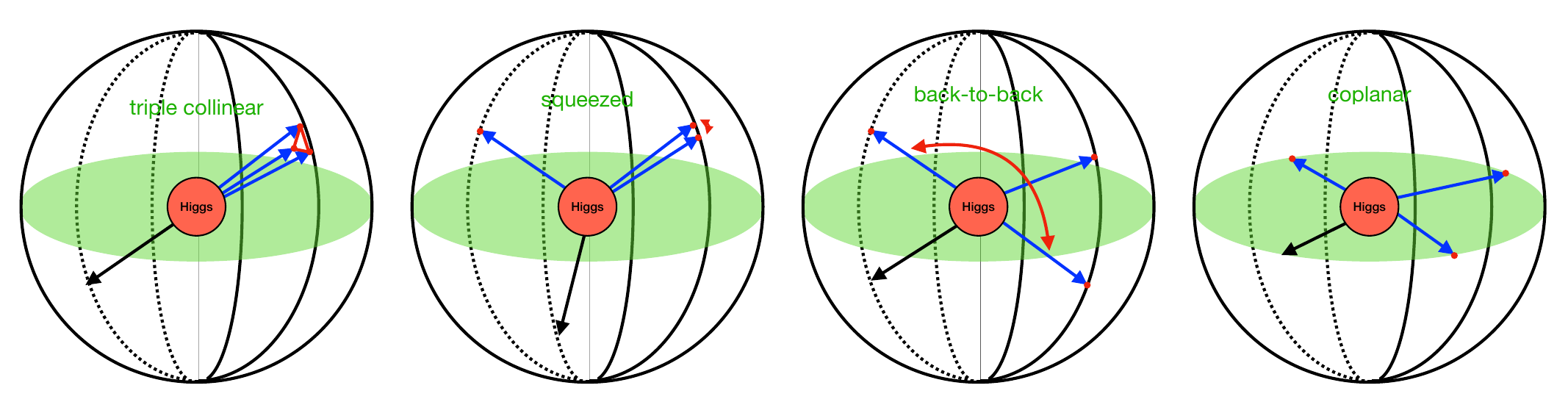}
    \caption{Different kinematic configurations for four final-state particles. The blue lines represent the particles that are measured by EEEC, and the black line is the fourth particle, whose direction is determined by momentum conservation. From left to right, we show triple-collinear, squeezed, back-to-back and coplanar limits.}
    \label{fig:configuration}
\end{figure*}

In this section, we discuss the kinematic limit of EEEC. We have already introduced some of the singular limits in the previous section when describing our analytic result. Here we summarize all possible interesting limits in the kinematic space in Tab.~\ref{tab:kinematic_limit}. The triple collinear limit corresponds to the case where three of the final-state particles are moving within a jet, and thus EEEC becomes a jet substructure observable in this limit. The coplanar limit refers to the configuration where three particles live in the same plane. While this is always true for three-particle final states, it becomes a kinematic constraint for four-particle final states and beyond. The squeezed (back-to-back) limit is reached when two final-state particles become collinear (back-to-back). We demonstrate these kinematic configurations for clarity in Fig.~\ref{fig:configuration}. It is also important to emphasize that all these kinematic limits are not cleanly separated. Instead, each of them overlaps with the others, as shown in Fig.~\ref{fig:kinematic_region}. Therefore, it is also interesting to look at a double kinematic limit, e.g. squeeze limit under the triple collinear limit.
\begin{table}
	\begin{center}
		\begingroup
		\renewcommand{\arraystretch}{1.5}
		\begin{tabular}{|c|c|c|}
			\hline
			Kinematic limit & $\{ x_1,\, x_2,\, x_3\}$ & $\{ s,\, \tau_1,\, \tau_2\}$ \\
			\hline
			Triple collinear & $x_{1,2,3}\to 0$ & $s\to 0$ \\
			\hline 
			Coplanar & $s_2\to 0$ &  $s\to 1$ \\
			\hline
			Squeezed & $x_1\to 0, x_2\sim x_3$ & $\tau_1\to 1$\\
			\hline
			Back-to-back & $x_1\to 1$ & $s\rightarrow 1 \,, \tau_1\rightarrow -1$ \\
			\hline
		\end{tabular}
		\endgroup
	\end{center}
	\caption{A summary of the EEEC kinematic limits. Each limit can be approached by either $\{x_1,x_2,x_3\}$ or $\{s,\tau_1,\tau_2\}$ variable set. The latter one reveals that the back-to-back limit is encompassed within the coplanar limit.}
	\label{tab:kinematic_limit}
\end{table}

\subsection{Triple collinear limit}
We start by comparing our expansion results in the triple collinear limit with the analytic results presented in Ref.~\cite{Chen:2019bpb}. To begin with, it's important to recall that the tree-level $n$-point energy correlator factorizes in the homogeneous collinear limit. This follows from the fact that $1\to (n+1)$ matrix elements factorize into the product of dijet matrix elements for $1\to 2$ processes and the $1\to n$ splitting functions. Moreover, the $(n+1)$-particle phase space decouples into the "hard" two-particle phase space and the collinear $n$-particle phase space. For the tree-level 3-point correlator, this factorization results in a dijet hard function and the so-called EEEC jet function in the triple collinear limit. Specifically, for gluon-initiated hadronic Higgs decays, the 3-point correlator gives rise to the EEEC gluon jet function. Similarly, the quark-initiated channel yields the EEEC quark jet function, which is identical to the one extracted from the process $e^+ e^- \to \text{hadrons}$~\cite{Yang:2022tgm, Chen:2019bpb}.

For convenience, we opt not to perform a full analytical expansion of our results in Eqs.~\eqref{eq:hgg_res_color} and \eqref{eq:hqq_res_color} in the triple-collinear limit. Instead, we randomly select 50 sets of $x_{1,2,3}$ values that are very close to $0$ and compare our complete results against the EEEC jet function presented in Ref.~\cite{Chen:2019bpb}. Across all randomly chosen sets, we observe good agreements. This serves as a robust validation of our analytic calculations, particularly given the non-trivial shape dependence of the jet function.

\subsection{Coplanar limit}

Secondly, we focus on the coplanar limit of EEEC. As mentioned above, the LO coplanar limit starts from the three-parton configuration at order $\alpha_s$, and our calculation actually gives the NLO coplanar limit at order $\alpha_s^2$.

As shown in the last two figures of Fig.~\ref{fig:hgghqqDsLogplot}, when approaching $s\rightarrow 1$, we find an edge $\phi_1+\phi_2=\pi$ that separates the EEEC distribution into two triangular regions, and the top one with $\phi_1+\phi_2>\pi$ exhibits more divergent behavior than the bottom one. Since the analytic result is expressed in terms of polylogarithms, there must exist some additional logarithmic divergence in the top triangular region. 

To understand the coplanar behavior at NLO, we first examine the coefficients for each basis in the function space. Only the coefficients for $g_6$, $g_7$, $g_8$, $g_9$, $g_{12}$, $g_{13}$ and $g_{17}$ start at $\mathcal{O}\left((1-s)^{-1}\right)$ when $s\rightarrow 1$. Then we expand these 7 bases in the coplanar limit, and it turns out that only $g_{17}$ gives rise to the logarithmic divergence. Explicitly, we find
\begin{align}
    g_{17}&\approx i \pi\, \theta\left(-\cos\frac{\phi_1}{2}\cos\frac{\phi_2}{2}\cos\frac{\phi_1+\phi_2}{2}\right) \nn\\
    &\times  \ln \left((1-s)^2 \tan^2\frac{\phi_1}{2}\tan^2\frac{\phi_2}{2}\tan^2\frac{\phi_1+\phi_2}{2}\right)\,.
\end{align}
The non-analytic $\theta$ function comes from the branch cut of $\text{Im}\left[\ln (2-x_1-x_2-x_3-s_2)\right]$ in Eq.~\eqref{eq:g17exp}. This logarithm is only nonzero when $2-x_1-x_2-x_3<0$, which simplifies into $\cos\frac{\phi_1}{2}\cos\frac{\phi_2}{2}\cos\frac{\phi_1+\phi_2}{2}<0$ using the parameterization described in Eqs.~\eqref{eq:xtoy}-\eqref{eq:ytotau}. This is precisely the top triangle region $\phi_1+\phi_2>\pi$ in the density figure. Regarding the other bases, they are less singular when $s\rightarrow 1$, so they only contribute to the power divergence $\frac{1}{1-s}$. In summary, we find for $H_{q\bar q}$ the following expression
\begin{align}
    \label{eq:Hqqcoplanar}
    H_{q\bar q}&\approx -\left(\frac{\alpha_s}{4\pi}\right)^2 \frac{\pi}{2}\, \theta\left(-\cos\frac{\phi_1}{2}\cos\frac{\phi_2}{2}\cos\frac{\phi_1+\phi_2}{2}\right)\nn\\
    &\times{\color{darkred} \frac{1}{1-s}\ln \left((1-s)^2 \tan^2\frac{\phi_1}{2}\tan^2\frac{\phi_2}{2}\tan^2\frac{\phi_1+\phi_2}{2}\right)}\nn\\
    &\times   f_{q\bar q}(\phi_1,\phi_2) +\text{ power divergence } +\mathcal{O}[(1-s)^{0}]
\end{align}
with the coefficient function
\begin{align}
	\label{eq:exp_fqq}
    f_{q\bar q}&=C_F(C_A+2C_F)\bigg[15+10\cos\phi_1+10\cos\phi_2\nn\\
    &+10\cos(\phi_1+\phi_2)+2\cos(\phi_1-\phi_2)-\cos(2\phi_1)\nn\\
    &-\cos(2\phi_2)-\cos(2\phi_1+2\phi_2)+2\cos(2\phi_1+\phi_2)\nn\\
    &+2\cos(\phi_1+2\phi_2)\bigg]\left(\sin\frac{\phi_1}{2} \sin\frac{\phi_2}{2} \sin\frac{\phi_1+\phi_2}{2}\right)^{-5}\,.
\end{align}
If we keep track of the color factors, we see that the logarithmic enhancement is from the partonic channel that involves gluons. This is expected because, from the perspective of EFT, soft gluons are allowed to slightly deviate from the plane, resulting in a soft enhancement proportional to $\ln(1-s)$. 

For the gluonic Higgs decay $H_{gg}$, we find the same structure but with a more complicated coefficient function $f_{gg}(\phi_1,\phi_2)$:
\begin{widetext}
\begin{align}
	f_{gg}&=-\frac{1}{32}(C_A+2C_F) T_F n_f\bigg[24 \cos \left(\phi _1\right)-22 \cos \left(2 \phi _1\right)+\cos \left(4 \phi _1\right)+4
	\cos \left(\phi _1-3 \phi _2\right)-4 \cos \left(\phi _1-2 \phi _2\right)\nn\\
	&\hspace{0.5cm}-24 \cos
	\left(\phi _1-\phi _2\right)+6 \cos \left(2 \left(\phi _1-\phi _2\right)\right)-4 \cos
	\left(2 \phi _1-\phi _2\right)+4 \cos \left(3 \phi _1-\phi _2\right)+24 \cos \left(\phi
	_2\right)-22 \cos \left(2 \phi _2\right)\nn\\
	&\hspace{0.5cm}+\cos \left(4 \phi _2\right)+24 \cos \left(\phi
	_1+\phi _2\right)-22 \cos \left(2 \left(\phi _1+\phi _2\right)\right)+\cos \left(4
	\left(\phi _1+\phi _2\right)\right)-24 \cos \left(2 \phi _1+\phi _2\right)\nn\\
	&\hspace{0.5cm}+6 \cos
	\left(2 \left(2 \phi _1+\phi _2\right)\right)-4 \cos \left(3 \phi _1+\phi _2\right)+4
	\cos \left(4 \phi _1+\phi _2\right)-24 \cos \left(\phi _1+2 \phi _2\right)+6 \cos
	\left(2 \left(\phi _1+2 \phi _2\right)\right)\nn\\
	&\hspace{0.5cm}-4 \cos \left(3 \phi _1+2 \phi _2\right)-4
	\cos \left(\phi _1+3 \phi _2\right)-4 \cos \left(2 \phi _1+3 \phi _2\right)+4 \cos
	\left(4 \phi _1+3 \phi _2\right)+4 \cos \left(\phi _1+4 \phi _2\right)\nn\\
	&\hspace{0.5cm}+4 \cos \left(3
	\phi _1+4 \phi _2\right)+45\bigg]\times \left(\sin\frac{\phi_1}{2} \sin\frac{\phi_2}{2} \sin\frac{\phi_1+\phi_2}{2}\right)^{-7}\nn\\
	&\hspace{0.5cm}+\frac{3}{32}C_A^2\bigg[-22 \cos \left(2 \phi _1\right)+\cos \left(4 \phi _1\right)+6 \cos \left(2 \left(\phi
   _1-\phi _2\right)\right)-22 \cos \left(2 \phi _2\right)-22
   \cos \left(2 \left(\phi _1+\phi _2\right)\right)+\cos \left(4 \phi _2\right)\nn\\
   &\hspace{0.5cm}+\cos \left(4 \left(\phi _1+\phi
   _2\right)\right)+6 \cos \left(2 \left(2 \phi _1+\phi _2\right)\right)+6 \cos \left(2
   \left(\phi _1+2 \phi _2\right)\right)+45\bigg]\times \left(\sin\frac{\phi_1}{2} \sin\frac{\phi_2}{2} \sin\frac{\phi_1+\phi_2}{2}\right)^{-7}\,.
\end{align}
\end{widetext}
More interestingly, the same logarithmic divergence exists in both $\mathcal{N}=4$ SYM and $e^+e^-$ annihilation. The $\mathcal{N}=4$ result is 
\begin{align}
	f_{\mathcal{N}=4}=48\times \left(\sin\frac{\phi_1}{2} \sin\frac{\phi_2}{2} \sin\frac{\phi_1+\phi_2}{2}\right)^{-3}\,,
\end{align}
where the normalization constant is slightly different from the result given in Ref.~\cite{Yan:2022cye}. For $e^+e^-$ annihilation, we find
\begin{align}
	f_{e^+e^-}&=-4C_F(C_A+2C_F)\bigg[2 \cos \left(\phi _2\right) \cos \left(\phi _1\right)\nn\\
 &\hspace{-0.5cm}+2 \cos \left(\phi _1+\phi _2\right)
	\cos \left(\phi _1\right)+2 \cos \left(\phi _1+2 \phi _2\right) \cos \left(\phi
	_1\right)\nn\\
 &\hspace{-0.5cm}+2 \cos \left(\phi _1\right)+\cos \left(2 \phi _1\right)+\cos \left(\phi_2\right)+\cos \left(\phi _1+\phi _2\right)\nn\\
 &\hspace{-0.5cm}+\cos \left(\phi _1+2 \phi _2\right)\bigg]\times \left(\sin\frac{\phi_1}{2} \sin\frac{\phi_2}{2} \sin\frac{\phi_1+\phi_2}{2}\right)^{-5}\,.
\end{align}
Notice that there are also logarithms involving $\phi_1$ and $\phi_2$ in Eq.~\eqref{eq:Hqqcoplanar}. This is precisely the overlap between the coplanar limit and the squeezed limit. Intuitively, under the coplanar configuration (as shown in the last picture of Fig.~\ref{fig:configuration}), two final-state particles are still allowed to be collinear with each other. 
For phenomenological applications, the existence of infrared divergence also suggests us to resum these logarithms to all orders. We defer this to future research.

\subsection{Squeezed limit}
In this subsection, we study the squeezed limit where two detectors are positioned atop each other while the third one is well separated from the first two. This corresponds to the limit of $x_1 \ll x_2 = x_3 = \eta$ and its permutations. Different permutations give the same result due to the bosonic symmetry of the three-point energy correlators. It is straightforward to extract the squeezed limit at the leading power, giving the following simple results:
\begin{align}
    H_{i}(x_1,x_2,x_3) \overset{x_1\rightarrow 0}{\underset{x_{2,3} = \eta}{\approx}} \left(\frac{\alpha_s}{4\pi}\right)^2 \frac{1}{4 \pi \sqrt{-s_2^2}} \left[\frac{ Q_i(\eta)}{x_1} + \mathcal{O}(x_1^0) \right]\,
\end{align}
with 
\begin{widetext}
    \begin{align}
    \label{eq:sQgg}
       Q_{gg}(\eta) &= n_f^2 T_F^2 \bigg[\frac{2
   \left(43 \eta ^4-1735 \eta ^3+7490 \eta ^2-10530 \eta
   +4740\right)}{45 (1-\eta ) \eta ^6}-\frac{8 \left(10 \eta
   ^3-70 \eta ^2+136 \eta -79\right) \ln (1-\eta )}{3 \eta
   ^7}\bigg]\nonumber\\
   &+C_F n_f T_F \bigg[\frac{12 \left(19 \eta
   ^2-62 \eta +47\right) \ln (1-\eta )}{\eta ^7}-\frac{21
   \eta ^4+185 \eta ^3-2530 \eta ^2+5130 \eta -2820}{5
   (1-\eta ) \eta ^6}\bigg]\nonumber\\
       &+C_A n_f T_F \bigg[\frac{8 \left(170 \eta ^4-1390 \eta ^3+3639
   \eta ^2-3869 \eta +1447\right) \ln (1-\eta )}{15 (1-\eta
   ) \eta ^7}  \nonumber\\
   &\hspace{2cm}+\frac{2 \left(609 \eta ^4-29905 \eta ^3+131210
   \eta ^2-188730 \eta +86820\right)}{225 (1-\eta ) \eta
   ^6}\bigg]\nonumber 
   \\
   & +C_A^2 \bigg[-\frac{8 \left(120 \eta ^4-840 \eta
   ^3+2482 \eta ^2-3092 \eta +1351\right) \ln (1-\eta )}{15
   (1-\eta ) \eta ^7}\nonumber\\
   &\hspace{2cm}-\frac{2 \left(1472 \eta ^4-17515 \eta
   ^3+83180 \eta ^2-144990 \eta +81060\right)}{225 (1-\eta )
   \eta ^6}\bigg]
    \end{align}
and
\begin{align}
\label{eq:sQqq}
   Q_{q\bar{q}}(\eta) &= C_A C_F \bigg[\frac{4 \left(40 \eta ^2-40 \eta +21\right)
   \ln (1-\eta )}{15 \eta ^6}-\frac{181 \eta ^3-678 \eta
   ^2+606 \eta -252}{45 (1-\eta ) \eta ^5}\bigg] \nonumber 
   \\
   & +C_F n_f T_F
   \bigg[\frac{4 \left(10 \eta ^2-10 \eta +3\right) \ln
   (1-\eta )}{15 \eta ^6}-\frac{43 \eta ^3-174 \eta ^2+138
   \eta -36}{45 (1-\eta ) \eta ^5}\bigg]\nonumber 
   \\
   &+C_F^2
   \bigg[-\frac{6 \left(19 \eta ^2-48 \eta +31\right) \ln
   (1-\eta )}{(1-\eta ) \eta ^6}-\frac{15 \eta ^3+64 \eta
   ^2-390 \eta +372}{2 (1-\eta ) \eta ^5}\bigg]\,,
\end{align}
\end{widetext}
where the highest power of $\eta$ in denominator is 7 for $Q_{gg}(\eta)$ and 6 for $Q_{q\bar{q}}(\eta)$. Interestingly, taking the $\mathcal{N}=1$ SYM limit by setting $T_F = 1/2,\,n_f = N_c,\, C_A = N_c,\, C_F = N_c$, the above two equations become identical,
\begin{multline}
    Q_{gg}(\eta)\big|_{\mathcal{N} =1} =  Q_{q\bar{q}}(\eta)\big|_{\mathcal{N} =1} =   \\
     N_c^2  \bigg[\frac{6 \left(2 \eta ^3+15 \eta ^2-45 \eta
   +30\right) \ln (1-\eta )}{(\eta -1) \eta ^6} 
   \\
   +\frac{3
   \left(4 \eta ^3+5 \eta ^2-60 \eta +60\right)}{(\eta -1)
   \eta ^5}\bigg]\,.
\end{multline}
However, the $\mathcal{N}=1$ limit of the corresponding result for $e^+ e^-$ annihilation gives a different result. Taking the collinear limit under the squeezed limit, i.e., $x_1\ll x_2 = x_3 = \eta \to 0$, the results in Eqs.~\eqref{eq:sQgg} and~\eqref{eq:sQqq} become
\begin{align}
\label{eq:csLimit}
  Q_{gg}(\eta)&  \approx \frac{1}{\eta} \bigg[ \frac{146 C_A n_f T_F}{225 }+\frac{872 C_A^2}{225 }+\frac{3
   C_F n_f T_F}{5 } \bigg] \,, \nonumber
   \\
    Q_{q\bar{q}}(\eta) & \approx \frac{1}{\eta} \bigg[ \frac{263 C_A C_F}{225}+\frac{59}{225} C_F n_f T_F+\frac{16
   C_F^2}{5}  \bigg]\,.
\end{align}
We find the above result for the quark-initiated channel in Higgs decay is identical to the corresponding result for $e^+ e^-$ annihilation as shown in~\cite{Yang:2022tgm}. It can be understood by the presence of common ingredients in the factorization theorem for the collinear limit under the squeezed limit as proposed in~\cite{Chen:2019bpb}. Taking the $\mathcal{N}=1$ SYM limit in Eq.~\eqref{eq:csLimit}, we obtained the following identical result
\begin{align}
    Q_{gg}(\eta)|_{\mathcal{N}=1} &\approx \frac{1}{\eta} N_c^2 \frac{9}{2} + \mathcal{O}(\eta)\,, \nonumber \\  
   Q_{q\bar{q}}(\eta)|_{\mathcal{N}=1} &\approx \frac{1}{\eta} N_c^2 \frac{9}{2} + \mathcal{O}(\eta) \,,
\end{align}
the simplicity of the above result serves as an additional indication of the correctness of our results for $H_{gg}(x_1,x_2,x_3)$ and $H_{q\bar{q}}(x_1,x_2,x_3)$.
\\
\subsection{Equilateral limit}

\begin{figure}[!htbp]
    \centering
    \includegraphics[scale=0.7]{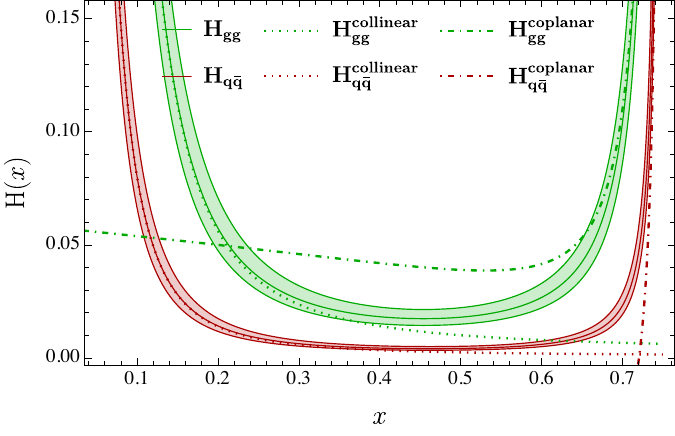}
    \caption{Equilateral EEEC, where all three angles are equal $x_1=x_2=x_3=x$. The fixed-order uncertainties are estimated in the same way as Fig.~\ref{fig:hggqqCsPlot}. The Solid lines correspond to the full result, the dotted lines represent its collinear expansion, and the dot-dashed lines depict the coplanar expansion.}
    \label{fig:equilateral}
\end{figure}

Lastly, we can also look at the configuration where all angles are the same $x_1=x_2=x_3=x$, namely equilateral EEEC. In this case, the phase space constraint from the Gram determinant in Eq.~\eqref{eq:delta4_fac} becomes $x<\frac{3}{4}$. The equilateral limit is different from other limits since it is not naturally a singular limit. Instead, it is a phase space cut imposed in the EEEC measurement function. In Fig.~\ref{fig:equilateral}, we plot both the gluon and quark distributions in the equilateral limit. They exhibit similar behavior to EEC, containing endpoint singularities on both sides. The only difference is that the $x\rightarrow \frac{3}{4}$ limit corresponds to the coplanar limit, instead of the back-to-back limit.

With the analytic result in hand, we can extract both collinear limit $x\rightarrow 0$ and coplanar limit $x\rightarrow \frac{3}{4}$ for equilateral EEEC. For collinear limit, we find $H_{i}(x)\overset{x\rightarrow 0}{\approx}\left(\frac{\alpha_s}{4\pi}\right)^2\frac{1}{4\sqrt{3}\pi} h_i^{\text{coll}}(x) $, where $h_{gg}^{\text{coll}}(x)$ and $h_{q\bar q}^{\text{coll}}(x)$ are presented in the following:
\begin{widetext}
\begin{align}
    &h_{gg}^{\text{coll}}(x)= C_A n_f T_F \left(\frac{\frac{1504 \kappa }{27
   \sqrt{3}}-\frac{4207}{135}}{x^3}+\frac{-\frac{5440 \kappa }{81 \sqrt{3}}+2 \pi
   ^2+\frac{136547}{5670}}{x^2}+\frac{-\frac{2624 \kappa }{81 \sqrt{3}}+\frac{10 \pi
   ^2}{3}-\frac{261893}{56700}}{x}-\frac{20896 \kappa }{729 \sqrt{3}}-\frac{34 \pi ^2}{3}+\frac{129746756}{893025}\right)\nn\\
   &+C_F
   n_f T_F \left(\frac{\frac{128 \kappa }{9
   \sqrt{3}}-\frac{344}{45}}{x^3}+\frac{\frac{64 \kappa }{27 \sqrt{3}}-4 \pi
   ^2+\frac{187213}{4725}}{x^2}+\frac{\frac{640 \kappa }{81 \sqrt{3}}+\frac{4 \pi
   ^2}{3}-\frac{430789}{28350}}{x}+\frac{2368 \kappa }{243 \sqrt{3}}+44 \pi ^2-\frac{259537657}{595350}\right)\nn\\
   &+ n_f^2 T_F^2\left(\frac{4}{35
   x^2}+\frac{11}{35 x}+\frac{191}{315}\right)+C_A^2
   \left(\frac{\frac{256 \kappa }{27 \sqrt{3}}+2 \pi
   ^2-\frac{2653}{135}}{x^3}+\frac{\frac{3008 \kappa }{81 \sqrt{3}}+\frac{4 \pi
   ^2}{3}-\frac{263512}{14175}}{x^2}+\frac{\frac{512 \kappa }{81 \sqrt{3}}-\frac{56 \pi
   ^2}{3}+\frac{24267847}{113400}}{x}\right.\nn\\
   &\left.-\frac{2368 \kappa }{729 \sqrt{3}}-\frac{2156 \pi ^2}{27}+\frac{1517131447}{1786050}\right)\,,
\end{align}
\begin{align}
    &h_{q\bar q}^{\text{coll}}(x)=C_A C_F \left(\frac{\frac{304 \kappa }{27 \sqrt{3}}-2 \pi
   ^2+\frac{779}{54}}{x^3}+\frac{\frac{3680 \kappa }{81 \sqrt{3}}-\frac{\pi
   ^2}{3}-\frac{32987}{1620}}{x^2}+\frac{\frac{416 \kappa }{27 \sqrt{3}}+\frac{49 \pi
   ^2}{3}-\frac{3121117}{18900}}{x}+\frac{4496 \kappa }{729 \sqrt{3}}+\frac{2255 \pi ^2}{27}-\frac{418459681}{510300}\right)\nn\\
   &+C_F n_f T_F
   \left(\frac{\frac{1856 \kappa }{27
   \sqrt{3}}-\frac{5354}{135}}{x^3}+\frac{\frac{14489}{405}-\frac{4736 \kappa }{81
   \sqrt{3}}}{x^2}+\frac{-\frac{640 \kappa }{27 \sqrt{3}}-8 \pi ^2+\frac{451462}{4725}}{x}-\frac{13568 \kappa }{729 \sqrt{3}}-\frac{160 \pi
   ^2}{3}+\frac{19740161}{36450}\right)\nn\\
   &+C_F^2 \left(\frac{-\frac{32
   \kappa }{27 \sqrt{3}}+4 \pi ^2-\frac{4543}{135}}{x^3}+\frac{-\frac{928 \kappa }{81 \sqrt{3}}+\frac{2
   \pi ^2}{3}+\frac{1277}{81}}{x^2}+\frac{-\frac{256 \kappa }{27 \sqrt{3}}-\frac{86 \pi
   ^2}{3}+\frac{3038537}{9450}}{x}-\frac{6976 \kappa }{729 \sqrt{3}}-\frac{3250 \pi ^2}{27}+\frac{45690151}{36450}\right)\,.
\end{align}
\end{widetext}
Here $\kappa\equiv\text{Cl}_2\left(\frac{i\pi}{3}\right)= \text{Im}\left[\text{Li}_2\left(e^{\frac{i\pi}{3}}\right)\right]$ is the Gieseking’s constant. Note that $\text{Cl}_2(\phi)\equiv-\int_{0}^{\phi} \log |2 \sin\frac{x}{2} | dx$ is the Clausen function. In both channels, the collinear asymptotics only exhibit power divergence, where $\frac{1}{x}$ leads to single logarithms upon integration over $x$, i.e., in the cumulant. This is expected because collinear EEEC is soft-safe and we should only observe single collinear logarithms in the cumulant at LO. On the other hand, we encounter both power and logarithmic divergences in the coplanar limit. The analytic expansion reads $H_{i}(x)\overset{x\rightarrow \frac{3}{4}}{\approx}\left(\frac{\alpha_s}{4\pi}\right)^2\frac{1}{4\pi} h_i^{\text{cop}}(x) $ with
\begin{widetext}
    \begin{align}
        &h_{gg}^{\text{cop}}(x)=C_A n_f T_F \left(\frac{-\frac{8192 \pi  \ln u}{6561}+\frac{299008 \pi }{19683}-\frac{32768 \pi  \ln 2}{6561}}{u}+\frac{-\frac{71680 \kappa }{729 \sqrt{3}}+\frac{328 \text{Li}_2(-3)}{243}-\frac{509972
   \pi ^2}{19683}+\frac{4890304}{25515}-\frac{7146688 \ln 2}{25515}}{\sqrt{u}}\right.\nn\\
   &\left.+\frac{360448 \pi  \ln u}{6561}+\frac{18464768 \pi }{19683}-\frac{32768}{243} \pi  \ln 3+\frac{1441792 \pi  \ln 2}{6561}\right)+C_F n_f T_F \left(\frac{-\frac{16384 \pi  \ln u}{6561}-\frac{8192 \pi }{2187}-\frac{65536 \pi  \ln 2}{6561}}{u}\right.\nn\\
   &\left.+\frac{\frac{12800 \kappa }{243 \sqrt{3}}-\frac{416
   \text{Li}_2(-3)}{243}+\frac{481808 \pi ^2}{19683}+\frac{5159168}{10935}-\frac{32681728 \ln 2}{32805}}{\sqrt{u}}-\frac{65536 \pi  \ln u}{6561}-\frac{1146880 \pi
   }{6561}+\frac{458752 \pi  \ln 3}{2187}-\frac{262144 \pi  \ln 2}{6561}\right)\nn\\
   &+n_f^2 T_F^2
   \left(\frac{16384 \pi }{19683 u}+\frac{\frac{2940928}{76545}-\frac{19693568 \ln 2}{229635}}{\sqrt{u}}+\frac{720896 \pi }{19683}\right)+C_A^2 \left(\frac{-\frac{57344 \pi  \ln u}{2187}-\frac{315392 \pi }{6561}-\frac{229376 \pi  \ln 2}{2187}}{u}\right.\nn\\
   &\left.+\frac{-\frac{60160 \kappa }{729 \sqrt{3}}-\frac{23860
   \text{Li}_2(-3)}{243}+\frac{599078 \pi ^2}{6561}+\frac{1273696}{76545}+\frac{246157024 \ln 2}{229635}}{\sqrt{u}}+\frac{622592 \pi  \ln u}{2187}-\frac{14434304 \pi
   }{6561}\right.\nn\\
   &\left.+\frac{1146880 \pi  \ln 3}{2187}+\frac{2490368 \pi  \ln 2}{2187}\right)\,,
    \end{align}
    \begin{align}
    &h_{q\bar q}^{\text{cop}}(x)=C_A C_F \left(\frac{-\frac{60160 \kappa }{729 \sqrt{3}}-\frac{13900 \text{Li}_2(-3)}{243}+\frac{20330 \pi ^2}{6561}+\frac{10180832}{25515}-\frac{49353632 \ln 2}{76545}}{\sqrt{u}}+\frac{-\frac{20480 \pi  \ln u}{2187}-\frac{112640 \pi }{6561}-\frac{81920 \pi  \ln 2}{2187}}{u}\right.\nn\\
    &\left.-\frac{4243456 \pi  \ln u}{6561}-\frac{5447680 \pi }{19683}-\frac{1982464 \pi  \ln 3}{2187}-\frac{16973824 \pi  \ln 2}{6561}\right)\nn\\
    &+C_F n_f T_F \left(\frac{-\frac{20480 \kappa }{729 \sqrt{3}}-\frac{2392 \text{Li}_2(-3)}{81}-\frac{76 \pi ^2}{243}-\frac{493888}{3645}-\frac{607424 \ln 2}{3645}}{\sqrt{u}}+\frac{40960 \pi }{6561 u}+\frac{8683520 \pi }{19683}\right)\nn\\
    &+C_F^2 \left(\frac{-\frac{6400 \kappa }{729 \sqrt{3}}-\frac{6704 \text{Li}_2(-3)}{243}+\frac{580264 \pi ^2}{6561}+\frac{14464}{729}+\frac{2488448 \ln 2}{2187}}{\sqrt{u}}+\frac{-\frac{40960 \pi  \ln u}{2187}-\frac{20480 \pi }{729}-\frac{163840 \pi  \ln 2}{2187}}{u}\right.\nn\\
    &\left.+\frac{6258688 \pi  \ln u}{6561}-\frac{3850240 \pi }{2187}+\frac{3211264 \pi  \ln 3}{2187}+\frac{25034752 \pi  \ln 2}{6561}\right)\,.
    \end{align}
\end{widetext}
In the above equations, we use $u\equiv \frac{3}{4}-x$ for convenience. Thus in the coplanar limit, we expect both collinear and soft divergences, which gives rise to $\frac{\ln u}{u}$ in the distribution and $\ln^2 (u)$ in the cumulant. Moreover, we observed a non-integer power $u^{-\frac{1}{2}}=(3-4x)^{-\frac{1}{2}}$ in the expansion. It will be interesting to understand the origin of this term. In Fig.~\ref{fig:equilateral}, we also plot the asymptotics for both channels, it is clear that the coplanar expansion deviates from the fixed-order earlier due to the aforementioned logarithmic divergences.

\section{Conclusion}
\label{sec:conclusion}

It is well known that high-precision QCD measurements require accurate calculations on the theoretical side. Despite significant progress in developing new techniques for multi-loop calculations, for example, Integration-by-part (IBP) reductions~\cite{CHETYRKIN1981159} and differential equations~\cite{KOTIKOV1991158,Gehrmann:2000zt}, very few of them can be directly applied to jet observables. This is mostly due to the complicated measurement functions associated with these observables, and thus one has to resort to computationally expensive Monte Carlo integration to obtain precise results with more digits. However, the numerical results become less reliable when the distribution contains large logarithms. In this sense, analytic calculations for jet observables are still necessary. Due to its simple measurement functions, the energy correlator stands out as a good candidate for developing techniques and achieving higher-order results.

In this paper, we perform the analytic calculations for LO EEEC in hadronic Higgs decays, specifically $H\rightarrow gg+X$ and $H\rightarrow q\bar q+X$. Instead of performing IBP reductions, we employ a good parameterization that allows the decomposition of the Gram determinant and facilitates the direct integration of the four-particle phase space. Topology identification is applied to simplify the integrand. The derived analytic results can be expressed using GPLs or classical polylogarithms up to transcendentality weight two. With widely used tools like \textsc{Mathematica}, our analytic findings can be efficiently evaluated on numerical points to high precision. It turns out that EEEC for $\mathcal{N}=4$ SYM, $e^+e^-$ annihilation, and Higgs decay share the same function space and analytic structure. Moreover, the gluonic channel $H_{gg}$ and quark channel $H_{q\bar q}$ exhibit very similar dependencies on $s$ and $\phi_{1,2}$. The universality suggests analytic structures of energy correlators that are independent of the underlying theories. The techniques developed here might also be applied to compute other phase space integrals or other observables.

We have already observed rich structures for three-point energy correlators at LO. While EEC is singular in the collinear and back-to-back limits, EEEC exhibits infrared divergences in multiple regions. As discussed in the previous section, these include triple-collinear limit, coplanar limit, squeezed limit, and others. The rich kinematics allow us to apply EEEC to jet productions at both $e^+e^-$ colliders and $pp$ colliders. 

Firstly, we can directly measure EEEC at $e^+e^-$ colliders. For LEP at $Q=91.2$ GeV, we are looking at the decays of photons or Z boson, while for future colliders like CEPC at $Q\approx 240$ GeV, we have access to the decay of Higgs boson. In this case, our analytic results provide the fixed-order theoretical data for the measurement of EEEC in hadronic Higgs decays.
The next step is to perform resummation for all possible large logarithms. As indicated in the kinematic analysis, a two-scale $\left(x_1/x_3,x_2/x_3\right)$ EFT is required to describe the singularities. This necessitates the derivation of a factorization theorem in every singular limit and the formulation of the resummed prediction for each of them. For the matching to the fixed-order result, we also need a profile scale that guides the transition from resummation to fixed-order behavior. For traditional event shapes like thrust~\cite{Becher:2008cf,Abbate:2010xh,Abbate:2012jh}, C-parameter~\cite{Hoang:2015hka}, heavy jet mass~\cite{Chien:2010kc}, one-dimensional quadratic profile functions are used to match dijet resummation to the fixed-order prediction. Since EEEC is a function of three variables, it could serve as a good candidate to study how to design such a profile scale in a high-dimensional distribution. For $H\rightarrow q\bar q$ channel, by incorporating renormalon subtraction~\cite{Schindler:2023cww} and accounting for non-perturbative power corrections, our result can be applied to precision Higgs measurements, including the determination of Yukawa couplings~\cite{Gao:2016jcm,Yan:2023xsd}. Regrading $H\to gg$ decay channel, which is likely to be observed at future lepton colliders, EEEC could also be a good observable to study gluon jets. In particular, comparing the theory prediction with the experimental data will provide us insights into gluon jet substructure and hadronization.


In addition, the collinear limit of EEEC serves as a valuable jet substructure observable at $pp$ colliders. By treating the collinear EEEC as the jet function in the EFT framework, we can focus on resumming the large logarithms in the collinear limit. This, combined with the hard function and jet algorithm, enables us to provide a theoretical prediction for jet production processes at $pp$ colliders. 
We can also establish connections between the EEEC calculations and other advancements in energy correlators. For example, the addition of labels for quarks and gluons in the calculation allows for the separation of different flavors, enabling the analytic result to be combined with track functions. Moreover, we can integrate our analytic result over two of the angles to obtain NLO projected energy correlators. The infrared divergences in this case will be canceled with the one-loop corrections for three-parton productions. All the analyses developed for projected energy correlators in 
$pp\rightarrow \text{jets}$ can then be generalized to Higgs decays.

\section*{Acknowledgements}
We thank Anjie Gao, Kyle Lee, Yibei Li, Ian Moult, Matthew D. Schwartz, Iain W. Stewart, Kai Yan and HuaXing Zhu for useful discussions on this work and related topics. We also thank HuaXing Zhu for useful comments on the manuscript. T.-Z Yang is supported by the European Research Council (ERC) under the European Union's Horizon 2020 research and innovation programme grant agreement 101019620 (ERC Advanced Grant TOPUP).

\newpage

\begin{widetext}
\appendix

\section{EEEC function space}

We provide the complete function space for Higgs EEEC. There are 7 weight-one bases $f_{1-7}$ as listed in Eq.~\eqref{eq:w1base} and 21 weight-two bases $g_{1-21}$ given below. Notice that we have already made the $S_3$ permutation symmetry explicit. With the function space, the analytic result for Higgs EEEC can be written as
\begin{equation}
    H_i=\sum_{i=1}^{7} r^{(1)}_i(x_1,x_2,x_3) f_i+\sum_{i=1}^{21} r^{(2)}_i(x_1,x_2,x_3) g_i
\end{equation}
with $r^{(1,2)}_i$ being the rational coefficients. The explicit forms of weight-two bases $g_{1-21}$ are given in the following:
\begin{align}
\label{eq:eeec_tw2_basis}
g_1&=\text{Li}_2\left(\frac{x_1}{x_1-1}\right),\quad g_2=\text{Li}_2\left(\frac{x_2}{x_2-1}\right),\quad g_3=\text{Li}_2\left(\frac{x_3}{x_3-1}\right),\notag\\
g_4&=2 \re\left[
   \text{Li}_2\left(\frac{s_2+x_1-x_2+2 x_2 x_3-x_3}{2
   \left(x_2-1\right) \left(x_3-1\right)}\right)-\text{Li}_2\left(\frac{2 x_1 x_2}{s_2-x_1+2 x_1
   x_2-x_2+x_3}\right)\right]+2
   \text{Li}_2\left(\frac{x_1}{x_1-1}\right)-2
   \text{Li}_2\left(\frac{x_3}{x_3-1}\right),\notag\\
g_5&=g_4(x_2\leftrightarrow x_3),\notag\\
g_6&=-2 i \im\left[\text{Li}_2\left(\frac{2 \left(x_1-1\right)
   x_2}{s_2-x_1+2 x_1 x_2-x_2+x_3}\right)\right]-2 i \im\left[\log
   \left(\frac{s_2-x_1+x_2-x_3}{2 x_1
   \left(x_2-1\right)}\right)\right] \re\left[\log \left(\frac{2
   \left(x_1-1\right) x_2}{s_2-x_1+2 x_1 x_2-x_2+x_3}\right)\right],\notag\\
g_7&=g_6(x_2\leftrightarrow x_3),\quad g_8=g_6(x_1,x_2,x_3\leftrightarrow x_2,x_3,x_1),\notag\\
g_9&=2 i \im\bigg[\text{Li}_2\left(\frac{s_2+x_1-x_2+
   x_3}{2(1- x_2)}\right)-\text{Li}_2\left(\frac{s_2+x_1-x_2-x_3}{2
   \left(x_1-1\right)}\right)-\text{Li}_2\left(\frac{2 x_1
   x_2}{s_2+x_1+x_2-x_3}\right)\notag\\
   &+\frac{1}{2} \log \left[\left(1-x_1\right)
   \left(x_2-1\right) \left(x_3-1\right)\right] \log
   \left(2-s_2-x_1-x_2-x_3\right)\bigg],\notag\\
g_{10}&=\pi^2,\quad g_{11}=-4\left[ \im\left[\log \left(2-s_2-x_1-x_2-x_3\right)\right]\right]^2,g_{12}=2 i \log \left(\frac{x_2\left(x_1-1\right)}{x_1
   \left(x_2-1\right)}\right) \im\left[\log
   \left(-s_2-x_1-x_2-x_3+2\right)\right],\notag\\
   g_{13}&=g_{12}(x_2\leftrightarrow x_3),\quad
g_{14}=\log \left(1-x_1\right) \log \left[\frac{\left(x_1-1\right)
   x_2}{x_1 \left(x_2-1\right)}\right]-\log \left(1-x_3\right) \log
   \left[\frac{x_2 \left(x_3-1\right)}{\left(x_2-1\right)
   x_3}\right],\notag\\
g_{15}&=g_{14}(x_2\leftrightarrow x_3),\quad g_{16}=g_{14}(x_1\leftrightarrow x_2),\notag\\
g_{17}&=-i
   \im\left[\text{Li}_2\left(\frac{s_2+x_1+x_2+x_3-2}{-s_2+x_1+x_2+
   x_3-2}\right)\right]-i \log
   \left[\frac{s_2^2}{\left(x_1-1\right) \left(x_2-1\right)
   \left(x_3-1\right)}\right] \im\left[\log
   \left(2-s_2-x_1-x_2-x_3\right)\right],\notag\\
g_{18}&=\text{Li}_2\left(\frac{x_1-x_2}{x_1(1- x_2)}\right)+\frac{1}{2}
   \log \left[\frac{x_2 \left(x_1-1\right) }{x_1
   \left(x_2-1\right)}\right] \log \left[\frac{x_3}{x_1(1-x_2)}\right],\notag\\
g_{19}&=g_{18}(x_2\leftrightarrow x_3), \quad g_{20}=g_{18}(x_1,x_2,x_3\leftrightarrow x_2,x_3,x_1),\notag\\
g_{21}&=-2 i
   \bigg\{\im\bigg[\text{Li}_2\left(\frac{s_1+x_1-x_2-x_3}{s_2+x_1-x_2-x_3}\right)-\text{Li}_2\left(\frac{s_1+x_1+x_2-x_3}{-s_2+x_1+x_2-x_3}\right)-\text{Li}_2\left(\frac{s_1+x_1+x_2-x_3}{s_2+x_1+x_2-x_
   3}\right)\notag\\
   &-\text{Li}_2\left(-\frac{s_1-x_1+x_2+x_3}{s_2+x_1-x_2-x_3}\right)-\text{Li}_2\left(\frac{2
   \left(x_1-1\right) x_2 \left(s_1+x_1-x_2+x_3\right)}{\left(s_2+x_1-2 x_1 x_2+x_2-x_3\right)
   \left(s_2-x_1+x_2+x_3\right)}\right)\notag\\
   &-\text{Li}_2\left(\frac{2 \left(x_1-1\right) x_2
   \left(s_1+x_1-x_2+x_3\right)}{\left(s_2+x_1-x_2-x_3\right) \left(s_2-x_1+2 x_1
   x_2-x_2+x_3\right)}\right)\bigg]+\im\left[\log \left(2-s_2-x_1-x_2-x_3\right)\right]\log
   \left(\frac{s_1-s_2}{s_1+s_2}\right)\notag\\
   & -\im\left[\log
   \left(\frac{s_2-s_1}{s_2-x_1+x_2+x_3}\right)\right] \re\left[\log
   \left(\frac{s_1+x_1-x_2-x_3}{s_2+x_1-x_2-x_3}\right)\right]\notag\\
   &-\im\left[\log
   \left(\frac{s_1+s_2}{s_2-x_1-x_2+x_3}\right)\right] \re\left[\log
   \left(\frac{s_1+x_1+x_2-x_3}{-s_2+x_1+x_2-x_3}\right)\right]\notag\\
   &-\im\left[\log
   \left(\frac{s_2-s_1}{s_2+x_1+x_2-x_3}\right)\right] \re\left[\log
   \left(\frac{s_1+x_1+x_2-x_3}{s_2+x_1+x_2-x_3}\right)\right]\notag\\
   &-\im\left[\log
   \left(\frac{s_1+s_2}{s_2+x_1-x_2-x_3}\right)\right] \re\left[\log
   \left(\frac{-s_1-x_1+x_2+x_3}{s_2-x_1+x_2+x_3}\right)\right]\notag\\
   &+\im\left[\log \left(\frac{2
   \left(s_1-s_2\right) \left(x_1-1\right) x_2}{\left(s_2+x_1-2 x_1 x_2+x_2-x_3\right)
   \left(s_2-x_1+x_2+x_3\right)}\right)\right]\notag\\
   &\times\re\left[\log \left(\frac{2 \left(x_1-1\right) x_2
   \left(-s_1+x_1-x_2+x_3\right)}{\left(s_2+x_1-2 x_1 x_2+x_2-x_3\right)
   \left(s_2-x_1+x_2+x_3\right)}\right)\right]\notag\\
   &+\im\left[\log \left(\frac{2 \left(s_1+s_2\right)
   \left(x_1-1\right) x_2}{\left(s_2+x_1-x_2-x_3\right) \left(s_2-x_1+2 x_1
   x_2-x_2+x_3\right)}\right)\right]\notag\\
   &\times \re\left[\log \left(\frac{2 \left(x_1-1\right) x_2
   \left(s_1+x_1-x_2+x_3\right)}{\left(s_2+x_1-2 x_1 x_2+x_2-x_3\right)
   \left(s_2-x_1+x_2+x_3\right)}\right)\right]\bigg\}\,.
\end{align}
We emphasize again that all bases are analytic, even though they are written in terms of real or imaginary parts of the logarithms. For convenience, we also list the expressions for the two square roots $s_{1,2}$ here:
\begin{equation}
    s_1\equiv\sqrt{x_1^2+x_2^2+x_3^2 - 2x_1x_2 -2x_1x_3-2x_2x_3},\quad s_2\equiv\sqrt{x_1^2+x_2^2+x_3^2 - 2x_1x_2 -2x_1x_3-2x_2x_3+4x_1x_2x_3}\,.
\end{equation}
A more compact form of the function space in terms of $\{s,\tau_1, \tau_2\}$ can be found in the $\mathcal{N}=4$ case in Ref.~\cite{Yan:2022cye}, which manifests the symmetries and is more convenient for theoretical analysis. The form presented here, however, is better suited for phenomenological applications.

 \end{widetext}
\bibliography{Higgs}
\end{document}